\documentclass[journal]{IEEEtran}
\usepackage{subcaption}
\usepackage{graphicx}
\usepackage{amsmath}
\usepackage{amssymb}
\usepackage{multirow}
\usepackage{cite}
\usepackage{url}
\usepackage{algorithmic}
\usepackage{algorithm}
\usepackage{xcolor}
\usepackage{xspace}
\usepackage{cases}
\usepackage{gensymb}
\usepackage[T1]{fontenc} 
\usepackage[cmintegrals]{newtxmath}
\usepackage{bm} 

\DeclareCaptionLabelSeparator{periodspace}{.\quad}
\begin{document}
\title{A Spherical Probability Distribution Model of the User-Induced Mobile Phone Orientation}

\author{Andr\'es Alay\'on Glazunov,~\IEEEmembership{Senior Member,~IEEE},   
        and Per Hjalmar Lehne 
\thanks{This work has been supported in part by a project within the VINNOVA funded Chalmers Antenna Systems VINN Excellence Centre (CHASE) at Chalmers, and by Telenor.

Andr\'{e}s Alay\'{o}n Glazunov is with the Department of Electrical Engineering, University of Twente, P.O. Box 217, 7500 AE Enschede, The Netherlands and he is also affiliated with the Department of Electrical Engineering (E2), Chalmers University of Technology, Gothenburg, Sweden (e-mails: a.alayonglazunov@utwente.nl, andres.glazunov@chalmers.se).
Per Hjalmar Lehne is with Telenor Research, Fornebu, Norway (e-mail: per-hjalmar.lehne@telenor.com)

Great part of this work was inspired by the late Prof. Per-Simon Kildal.}}
\maketitle
\begin{abstract}
This paper presents a statistical modeling approach of the real-life user-induced randomness due to mobile phone orientations for different phone usage types. As well-known, the radiated performance of a wireless device depends on its orientation and position relative to the user. Therefore, realistic handset usage models will lead to more accurate Over-The-Air characterization measurements for antennas and wireless devices in general. We introduce a phone usage classification based on the network access modes, e.g., voice (circuit switched) or non-voice (packet switched) services, and the use of accessories such as wired or Bluetooth handsets, or a speaker-phone during the network access session. The random phone orientation is then modelled by the spherical von Mises-Fisher distribution for each of the identified phone usage types. A finite mixture model based on the individual probability distribution functions and heuristic weights is also presented. The models are based on data collected from built-in accelerometer measurements. Our approach offers a straightforward modeling of the user-induced random orientation for different phone usage types. The models can be used in the design of better handsets and antenna systems as well as for the design and optimization of wireless networks.
\end{abstract}
\IEEEpeerreviewmaketitle
\section{Introduction}\label{sec:1}
User-induced randomness of the phone orientation has been known to influence the performance of wireless devices, such as mobile phones. User-induced randomness is understood here as the variation of the usage positions of wireless devices in real-life situations. Currently, fixed usage type positions are largely based on the hypothesis that certain phone orientations should dominate depending on the services accessed. However, they do not fully describe the actual span of possible usage positions of a mobile phone. These are needed for devising realistic Over-The-Air (OTA) performance testing of wireless devices capable of providing satisfactory Quality-of-Service (QoS) to end users \cite{MimoOtaTest_GlazKolLait12}.

The device performance is sensitive to the propagation environment which in turn depends, among other things, on the usage mode of the device, e.g., device orientation in voice or non-voice modes ~\cite{CorrCapMIMO_Kildal04,OTATestingMultiPath_Kildal12}. In OTA testing, two edge environments can be defined: the RIMP (Rich Isotropic MultiPath) environment and the Random-LOS (Random Line-OF-Sight) environment. In RIMP, the transmit signal reaches the receive antenna isotropically through various propagation paths. On the other hand, in Random-LOS there is only one path, i.e., the direct signal path between the transmit and receive antennas that is subjected to user-induced randomness. In RIMP, performance is independent from the device orientation; yet, it depends on the device usage, e.g, power absorption, impedance mismatch/detuning or both, as a result of the user body and different usage positions. In Random-LOS, randomness is induced mainly by the usage position and orientation relative the base station. This results in a random Angle-of-Arrival (AoA) and polarization of the incoming waves. Hence, the characterizations of wireless devices can be readily done in terms of polarization deficiencies of antennas and their impact on throughput performance as shown in \cite{ARazavi2016x,ARazavi2017}. The OTA performance in RIMP can be tested in reverberation chambers. On the other hand, OTA performance in ``pure-LOS" has been traditionally characterized in anechoic chambers. Hence, performance in Random-LOS can be measured in anechoic chambers too.

It has already been demonstrated that random device orientation affects the propagation characteristics~\cite{ImpactAntenna_Mellios_12_6348849}. In the recently introduced Random-LOS OTA characterization method, which uses the eponymous channel model, the effect of user-induced randomness on the system performance is naturally integrated~\cite{RethinkingWireless_Kildal_13,NewApproachToOTA_Kildal_13,CostEffectMeasur_Kildal_13}. Initial work pertaining the characterization of the user-induced randomness has been presented in \cite{UserRandomNess_Lehne_15,UserRandomNess_Lehne_16}. However, currently available OTA tests do not take into account the user-induced randomness, but only fixed usage types, due to non-existing probabilistic usage models of modern versatile (smart) mobile phones.

In order to fill the aforementioned knowledge gap we present here the first systematic model addressing the user-induced mobile phone orientation randomness for voice and non-voice applications. The relevance of making such a model follows from the fact, among other things, that there is a need to produce OTA tests of antennas and wireless devices that incorporate actual user orientation in a realistic and relevant manner. The scope of this paper is therefore to use collected accelerometer sensor data from smart phones and analyze them in order to suggest a proper model for the device orientation \cite{UserRandomNess_Lehne_16}. The underlying method of collecting data from smart phones for this purpose, can also be adapted for use in, e.g., network optimization to improve use experience based on knowledge about usage modes, however this is outside the scope of the current work.

The contributions of this paper are summarized as follows:
\begin{itemize}
  \item We suggest an approach to extract relevant data from accelerometer measurements obtained with Android phones by combining outlier removal with duplicate data removal.
  \item We introduce a usage type classification that follows a user's access to network services in terms of the two possible modes: voice (meaning circuit switched) or non-voice (meaning packet switched) and with the corresponding use of speaker and wired- or Bluetooth-headset following the user's natural behaviour. That is no pre-defined usage type positions are investigated, but the focus is on a users's real-life handling of the phones.
  \item We propose the von Mises-Fisher (vMF) distribution, i.e., the equivalent of the Gaussian distribution defined on the unit sphere \cite{mardia2009directional} to model the statistical probability distribution function of the orientation of  mobile phones for each specific usage type.
  \item An easy-to-use finite mixture model (FMM) is presented based on individual vMF distributions and on heuristic mixing probabilities corresponding to different phone usage types. The provided vMF distributions as well as the vMF FMM can be used to reproduce the desired random realizations of the user-induced orientations of the mobile phones and used as an input to further analysis, e.g., OTA characterization of mobile phones, or wireless network simulations.
  \item The extracted model parameters corresponding to the phone usage types classification are mapped to the classification of a phone's orientation in spherical coordinates given in \cite{UserRandomNess_Lehne_16}. Then, in order to harmonize our study with the existing fixed usage type definitions, e.g., defined by 3GPP we infer corresponding usage types from the analyzed accelerometer data. In this way, the provided model becomes more useful for the objective of standardized OTA methods.
\end{itemize}

The remainder of the paper is organized as follows: In Section~\ref{sec:2} we present a description of the data collection and classification corresponding to different usage of a mobile phone for voice and non-voice service types, here we also present a straightforward method to select data representative of user-induced randomness. In Section~\ref{sec:3} we introduce the von Mises-Fisher directional distribution model of the device orientation and validation procedure based on Quantile-Quantile plots (Q-Q plots). Section~\ref{sec:4} presents an analysis of the modelling results. Also here we suggest a finite mixture model for the identified phone usage types. A mapping of phone usage types to typical usage positions, assumed by 3GPP, is provided too. Conclusions and future work are outlined in  Section~\ref{sec:5}.
\begin{table}[!t]
\renewcommand{\arraystretch}{1.3}
\caption{Binary category classification of accelerometer data used to define phone usage type.
}
\centering
\setlength{\tabcolsep}{3pt}
\begin{tabular}{c |c c@{}}
    \hline \hline
    Bit                  & $0$ & $1$    \\
    \hline
    Service              & VOICE   & NON-VOICE       \\
    Wired-Headset        & NO      & YES        \\
    Speaker-Phone        & OFF     & ON         \\
    Bluetooth-Headset    & OFF     & ON         \\
    \hline \hline
\end{tabular}\label{tab:1}
\end{table}
\section{Acceleration data collection and interpretation}\label{sec:2}
\subsection{Measurements and data classification}
\begin{figure}[!t]
    \centering
    \begin{minipage}{0.5\textwidth}
        \centering
        \includegraphics[width=0.5\linewidth]{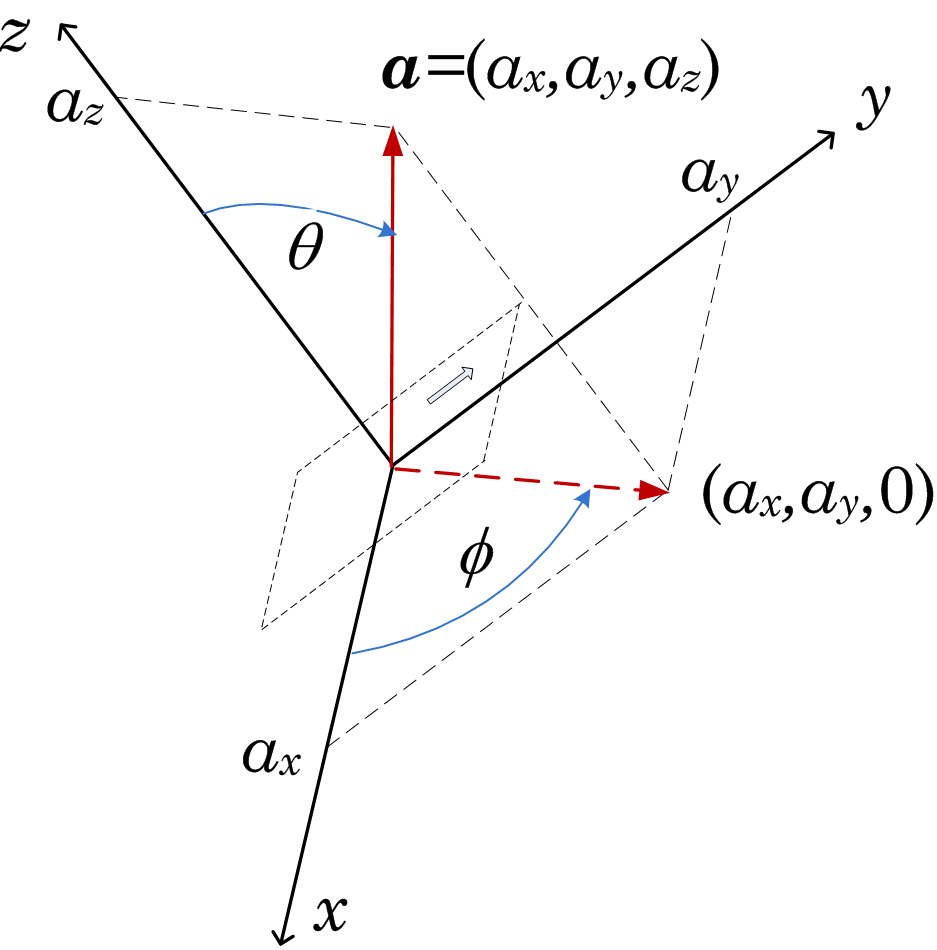}
        \caption{Gravity acceleration vector in the device coordinate system.}
        \label{figure:coordinate}
    \end{minipage}
\end{figure}

Modern smart phones offer ample opportunities in their sensor capabilities. In this paper we use the built-in 3D linear acceleration sensors (i.e., the accelerometer) to determine the device orientation. The acceleration vector components are given in the device local coordinate system and delivers values in m/s\textsuperscript{2}. The measured values include the gravity acceleration vector. If the phone is stationary the length of the vector is $g=9.8 \mathrm{m/s}^2$ (or other more exact value that depends on the actual geographical location). The device local coordinate system definition for Android phones is shown in Fig.~\ref{figure:coordinate} \cite{DeviceCoordinates_Android}, where the positive direction is defined as the direction of increased acceleration. This means that the gravity acceleration vector is pointing away from the Earth's centre.

\begin{table}[!]
\renewcommand{\arraystretch}{1.3}
\caption{Estimated parameters corresponding to the von Mises-Fisher directional probability distribution function, i.e., the concentration parameter $\hat{\kappa}$, and the mean direction vector $\boldsymbol{\hat{\mu}}$ in Cartesian co-ordinates and the corresponding orientation angles in spherical coordinates, $\hat{\phi}_{\mu}$ and $\hat{\theta}_{\mu}$. Heuristic finite mixture model (FMM) weights $\pi_{ijkl}$ and the number of samples $N_{\mathrm{s},ijkl}$ used to compute them.
}
\centering
\setlength{\tabcolsep}{3pt}
\begin{tabular}{c |c c c c c c|c c c@{}}
    \hline \hline
     Ph. Usage&  \multicolumn{6}{c}{vMF Distribution Parameters} &  \multicolumn{2}{|c}{FMM}   \\ 
    \hline \hline
    $\{ijkl\}$&  $\hat{\kappa}$  &  $\hat{\mu}_x$ &  $\hat{\mu}_y$ & $\hat{\mu}_z$  &  $\hat{\phi}_{\mu}, [^{\circ}]$ & $\hat{\theta}_{\mu}, [^{\circ}]$ & $\pi_{ijkl}$  & $N_{\mathrm{s},ijkl}$\\
    \hline \hline
    $0000$  &  $3.23$  &  $ 0.27$ & $ 0.93$  &  $ 0.24$  &  $ 73.97$  & $76.37$ &  $0.5433$ & 47988\\
    $0001$  &  $1.88$  &  $ 0.87$ & $ 0.50$  &  $-0.01$  &  $ 29.88$  & $90.75$ &  $0.0494$ & 4365\\
    $0010$  &  $4.17$  &  $ 0.23$ & $ 0.82$  &  $ 0.52$  &  $ 74.01$  & $58.51$ &  $0.0262$ & 2312\\
    $0100$  &  $2.10$  &  $ 0.08$ & $ 0.30$  &  $ 0.95$  &  $ 75.69$  & $17.85$ &  $0.2253$ & 19903\\
    \hline \hline
    $1000$  &  $1.37$  &  $-0.04$ & $ 0.65$  &  $ 0.76$  &  $ 93.73$  & $40.85$ &  $0.0694$ & 6134\\
    $1010$  &  $4.99$  &  $-0.06$ & $ 0.88$  &  $ 0.47$  &  $ 93.78$  & $62.01$ &  $0.0365$ & 3221\\
    $1100$  &  $3.39$  &  $ 0.08$ & $ 0.80$  &  $ 0.59$  &  $ 84.44$  & $53.73$ &  $0.0499$ & 4405\\
    \hline \hline
\end{tabular}\label{tab:2}
\end{table}

A smart phone application (app) has been installed onto a number of phones, and the data has been uploaded to a server which aggregates the data automatically into a searchable database~\cite{UserRandomNess_Lehne_15}.  This app records sensor values from the phone in the background while the smart phone is active, i.e., during a phone call or non-voice session. The app collects samples approximately once per second. The phones belonged to ordinary users, and no instruction of specific behaviour was given, since we wanted to record the normal usage. Once installed and registered, the app was running as a background service on the phone, and was not interacting in any way with the user. Hence, during a service session the user behaves naturally, i.e., carries out with her common behaviour while using the mobile phone.

The data is further classified into four binary and self-explanatory categories as shown in Table.~\ref{tab:1}. The voice service in this context means circuit switched (CS) voice, not VoIP services like Skype or Viber which belong to the non-voice services. The value $0$ is prescribed to the voice service, and to the NO and OFF specifications, while at the same time prescribing the value $1$ to the non-voice service, and to the YES and ON specifications we get a $4-$bit binary representation of the data giving $16$ possibilities in total, out of which only some make practical sense. For example, a user that during voice service doesn't use a wired-headset, a speaker phone or a Bluetooth-headset can be denoted as \emph{phone usage type: $0000$.} Another user, that during non-voice service, uses a wired-headset, a speaker phone and a Bluetooth-headset can be denoted as \emph{phone usage type: $1111$.} The likelihood of observing phone usage type $0000$ is high since it corresponds to the traditional way to talk over the phone. On the other hand phone usage type $1111$ doesn't really make sense in a practical scenario.

\subsection{Acceleration data modelling and estimates}
The gravity acceleration vector is used to estimate the accelerometer orientation \cite{Mizell03}. The collected acceleration data is modelled by two terms
\begin{equation}\label{eq:1}
  \boldsymbol{a}=\boldsymbol{Rg}+\boldsymbol{a}_\mathrm{e},
\end{equation}
where each data sample $i$ is actually a 3D vector $\boldsymbol{a}_i=(a_{xi},a_{yi},a_{zi})$, $\boldsymbol{g}$ has the same magnitude as the gravity acceleration vector, but is directed in the opposite direction, i.e., upwards. If there are no other forces than gravity acting upon the device, i.e., the device is still or moving at a uniform rectilinear velocity, then $\boldsymbol{a}_\mathrm{e}=0$. In this case, the measured acceleration is just the projection of the vector $\boldsymbol{g}$ into the local coordinate system of the device. Hence, the change of direction of the vector $\boldsymbol{g}$ is determined by the rotation matrix $\boldsymbol{R}$, such that $\|\boldsymbol{Rg}\|=g$, where the symbol $\|\boldsymbol{x}\|$ denotes the Euclidean norm of vector $\boldsymbol{x}$. Additive errors are modelled by $\boldsymbol{a}_\mathrm{e}$ since the actual non-inertial movement pattern (due to acceleration or deceleration) of the user is not known. The true distribution of $\boldsymbol{a}_\mathrm{e}$ is not known either nor are the distributions of the elements of the rotation matrix $\boldsymbol{R}$.

\begin{figure}[!t]
\centering
\scalebox{0.24}{\includegraphics[trim = 90 0 100 0,clip]{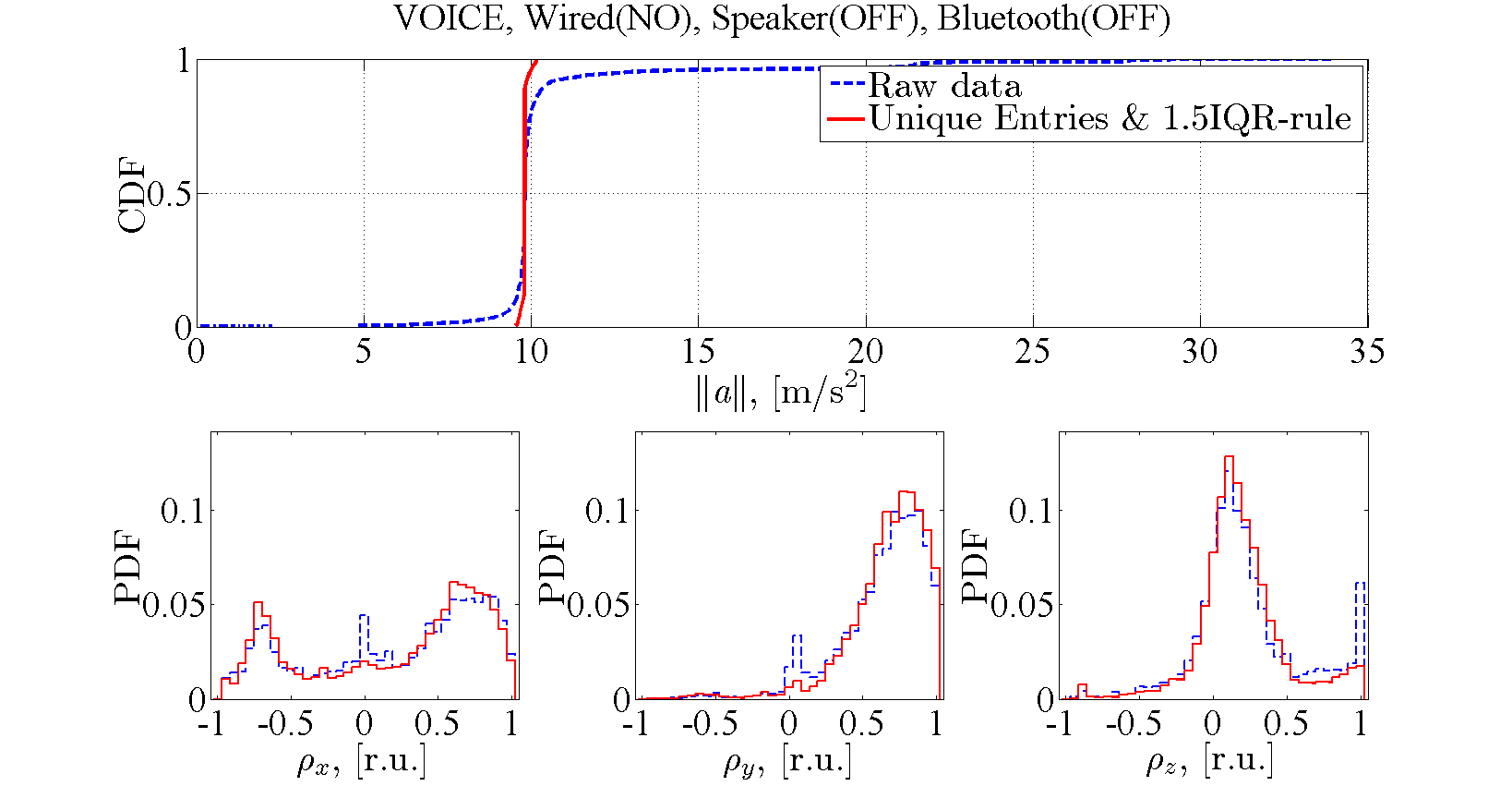}}%
\caption{Example of empirical Cumulative Distribution Function of the magnitude of the measured acceleration $\|\boldsymbol{a}\|$ (upper subplot) and the corresponding empirical Probability Distribution Function of the three cartesian components of the orientation vector $\boldsymbol{\rho}$. Plots corresponding to the raw data and to data after removing outliers are shown for voice service users that used no wired-headset, no speaker-phone and no Bluetooth-headset, i.e. usage type:$0000$.}
\label{fig:2}
\vspace{0em}
\end{figure}

The above physical arguments lead us to the following observations: first, the acceleration data samples can be considered \emph{random}; secondly, there may be \emph{outliers} (i.e., samples that are much larger or smaller than the rest of the samples), and thirdly, ambiguity in acceleration data samples can be reduced by choosing those which magnitude is close to the magnitude of the \emph{gravity acceleration vector}, i.e., $\|\boldsymbol{a}\|\approx g$. Following these observations we seek to remove outliers from data samples that are much larger or much smaller than $g$.
To identify and remove outliers from the data we use the \emph{$1.5$IQR-rule}. A data sample $\boldsymbol{a}_i$ is considered an outlier if $\|\boldsymbol{a}_i\|>Q_3+1.5IQR$ or $\|\boldsymbol{a}_i\|<Q_1-1.5IQR$, where $IQR=Q_3-Q_1$ is the Inter-Quartile Range, $Q_1$ and $Q_3$ are the $25\%-$ and the $75\%-$quartiles, respectively. We use the fact that the size of the $IQR$ is an indication of the spread of the middle half of the data. In our case it should be concentrated in a very narrow region around $Q_2$, the $50\%-$quartile, i.e., the median.

Removing the outliers will have a filtering effect on the measurement data, which is a desirable effect since it improves the quality of the data. Other errors, e.g., due to hardware impairments, noise etc., are not considered here. Their importance will become apparent only for very accurate estimations of the user location \cite{SensorReliability_Blum_13,ExperimentPhoneSensors_Zhizhong_13_6621233}.

Another aspect related to the data quality or the number of useful samples, i.e., measurements coming from completely random user positions. For example, in realistic situations, it is unlikely to observe a large number of identical consecutive acceleration values. This is because, the movement pattern of a user is not completely stationary. Stationarity of users is outside the analysis presented here. However, we suggest, as a first step toward reducing measurement artifacts to take into account only \emph{unique acceleration samples}. In this way, we remove duplicate data that cannot be clearly understood or that is not representative for the general usage patterns of mobile phones. In the future, as the size of data base increases, especially with measurements coming from a large number of different users, more sophisticated data processing will become necessary.

Hence, in this paper we remove data artifacts following two criteria: a data sample must not be an outlier, i.e., the acceleration vector can't be much larger or smaller than $g$ and it's components must be unique. Next we present a statistical distribution model for user-induced mobile phone orientation.

\begin{figure}[!ht]
\centering
\begin{subfigure}{\columnwidth}
\centering
\scalebox{0.9}{\includegraphics[width=\textwidth]{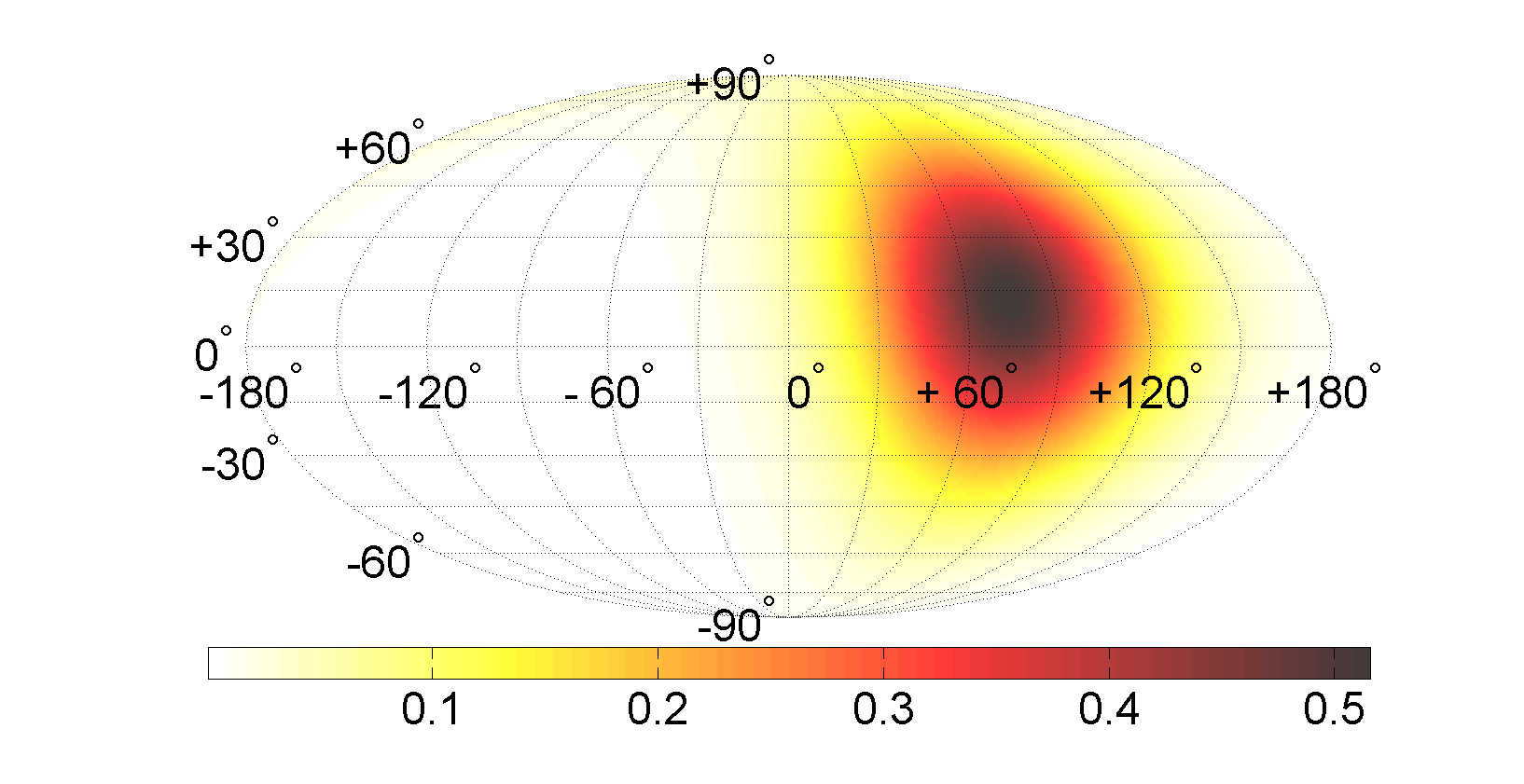}}%
\caption{}%
\label{fig:3a}
\end{subfigure}~
\\
\begin{subfigure}{\columnwidth}
\centering
\scalebox{0.9}{\includegraphics[width=\textwidth]{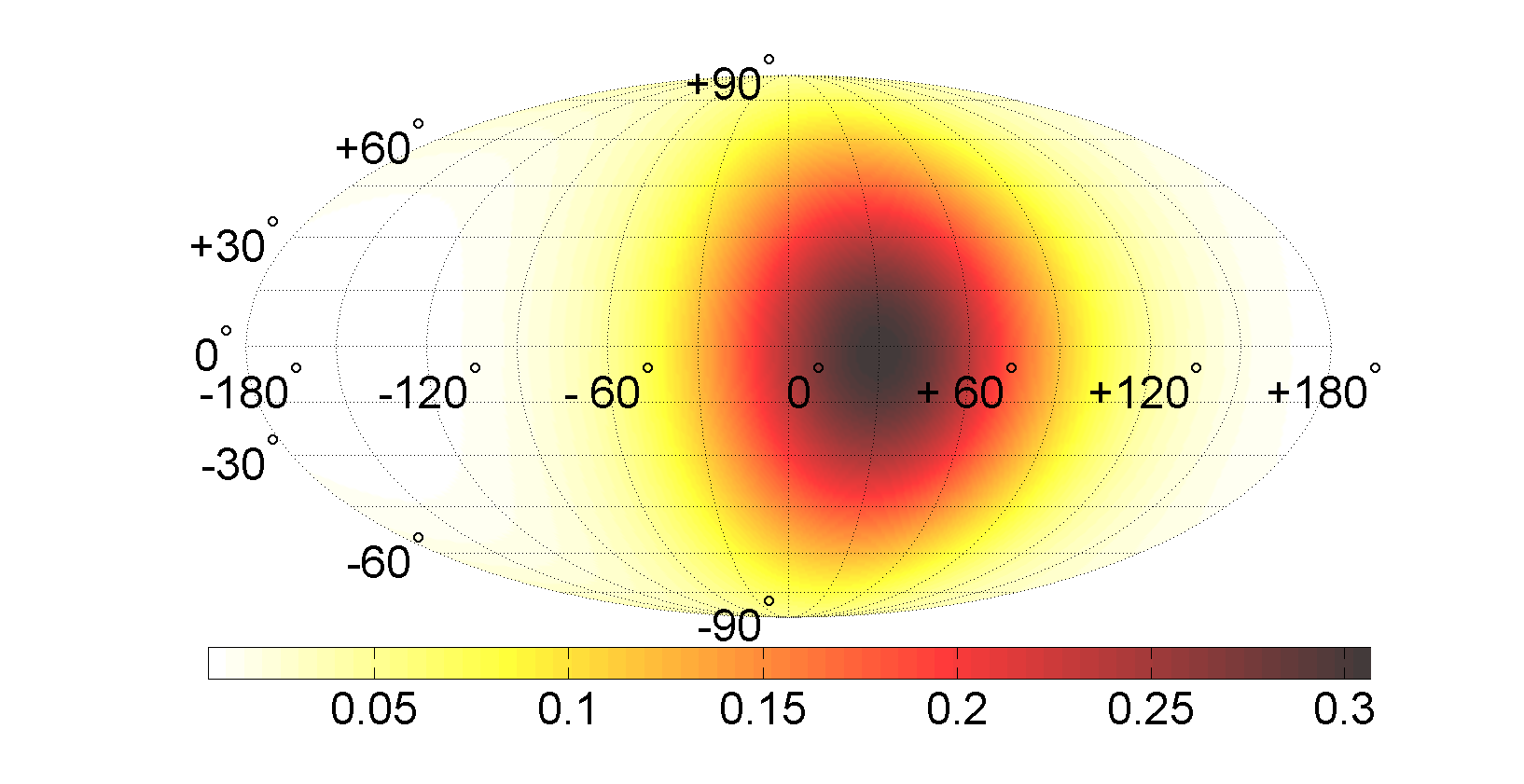}}%
\caption{}%
\label{fig:3b}
\end{subfigure}~
\\
\begin{subfigure}{1\columnwidth}
\centering
\scalebox{0.9}{\includegraphics[width=\textwidth]{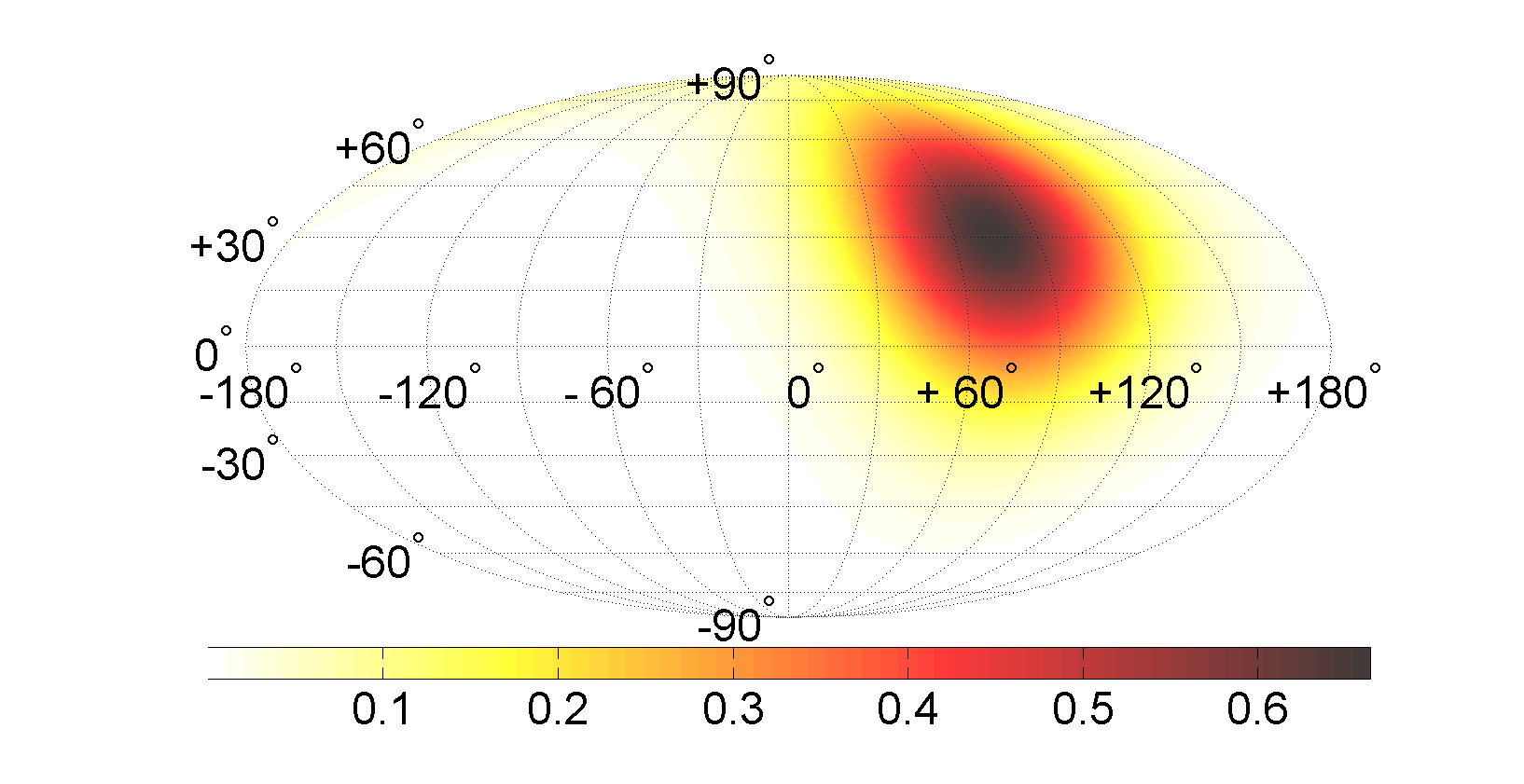}}%
\caption{}%
\label{fig:3c}%
\end{subfigure}~%
\\
\begin{subfigure}{\columnwidth}
\centering
\scalebox{0.9}{\includegraphics[width=\textwidth]{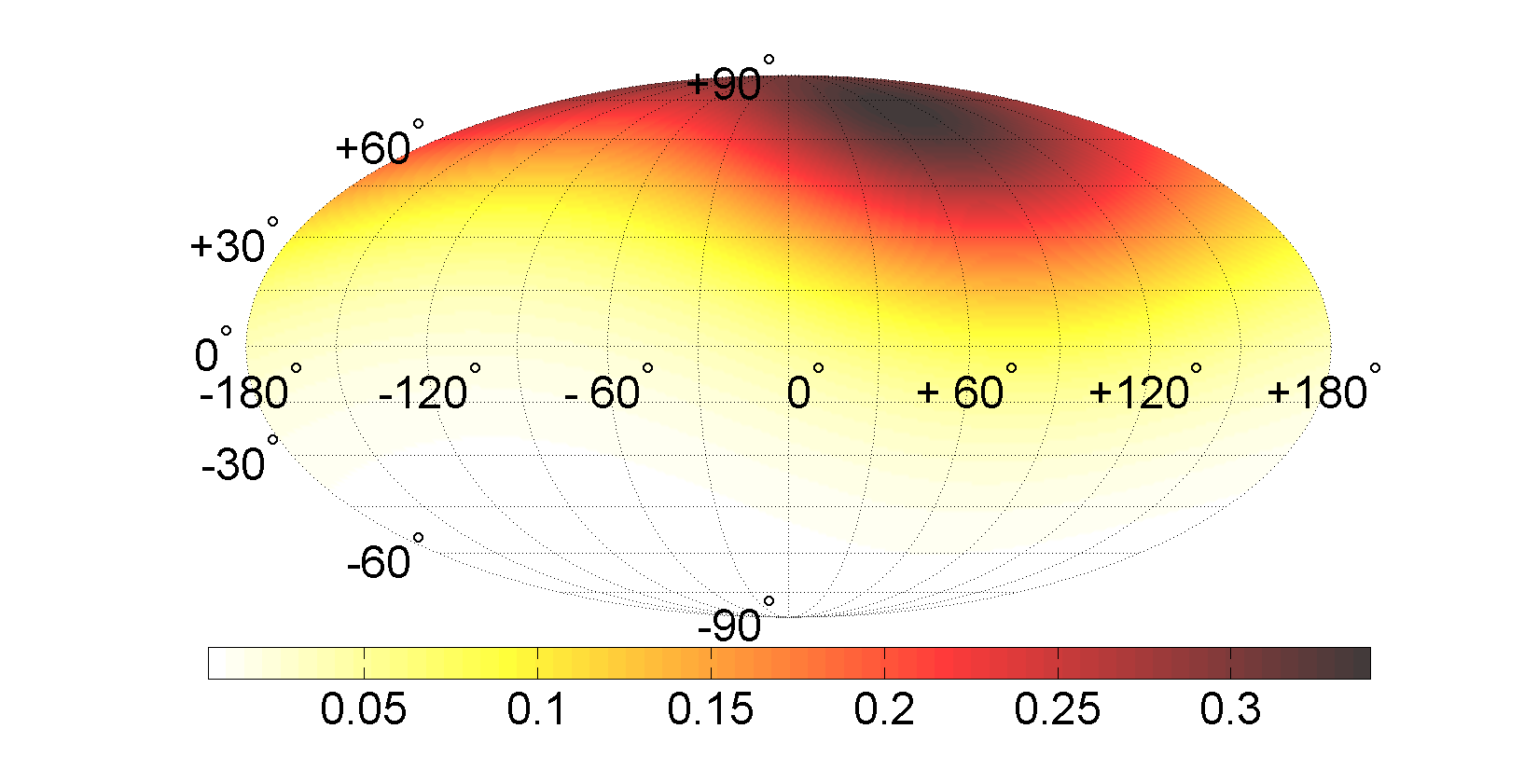}}%
\caption{}%
\label{fig:3d}%
\end{subfigure}%
\vspace{-0.5em}
\caption{3D von Mises-Fisher probability distribution function represented on a 2D Mollweide map projection for voice users, more specifically for phone usage types;  a) $0000$, b) $0001$, c) $0010$ and d) $0100$. See Table.~\ref{tab:1} and Table.~\ref{tab:2} for phone usage type and distribution parameter specification, respectively.}
\vspace{-0.5em}
\label{fig:3}
\end{figure}

\begin{figure}[!ht]%
\centering
\begin{subfigure}{\columnwidth}
\centering
\scalebox{0.9}{\includegraphics[width=\textwidth]{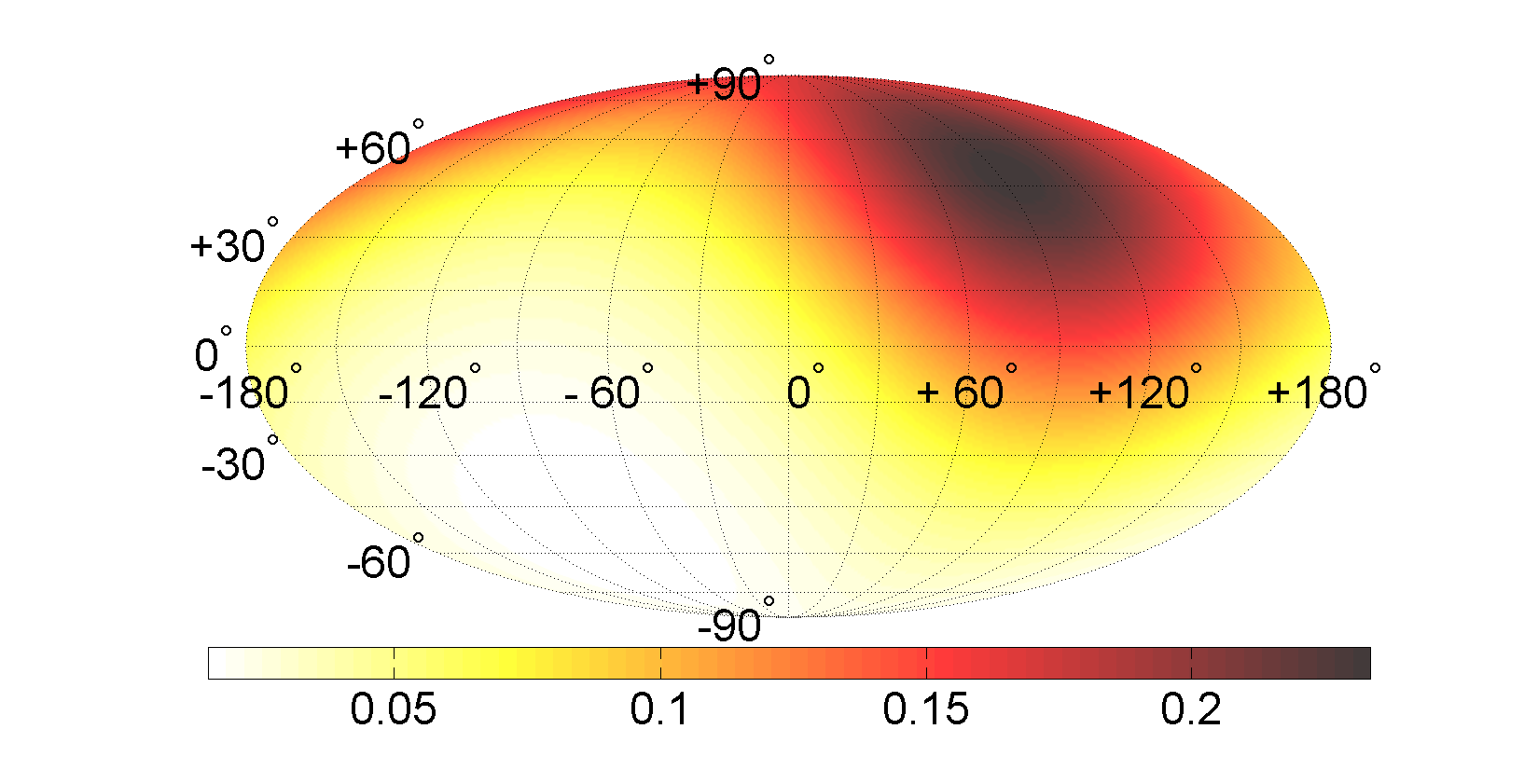}}%
\caption{}%
\label{fig:4a}%
\end{subfigure}~
\\
\begin{subfigure}{\columnwidth}
\centering
\scalebox{0.9}{\includegraphics[width=\textwidth]{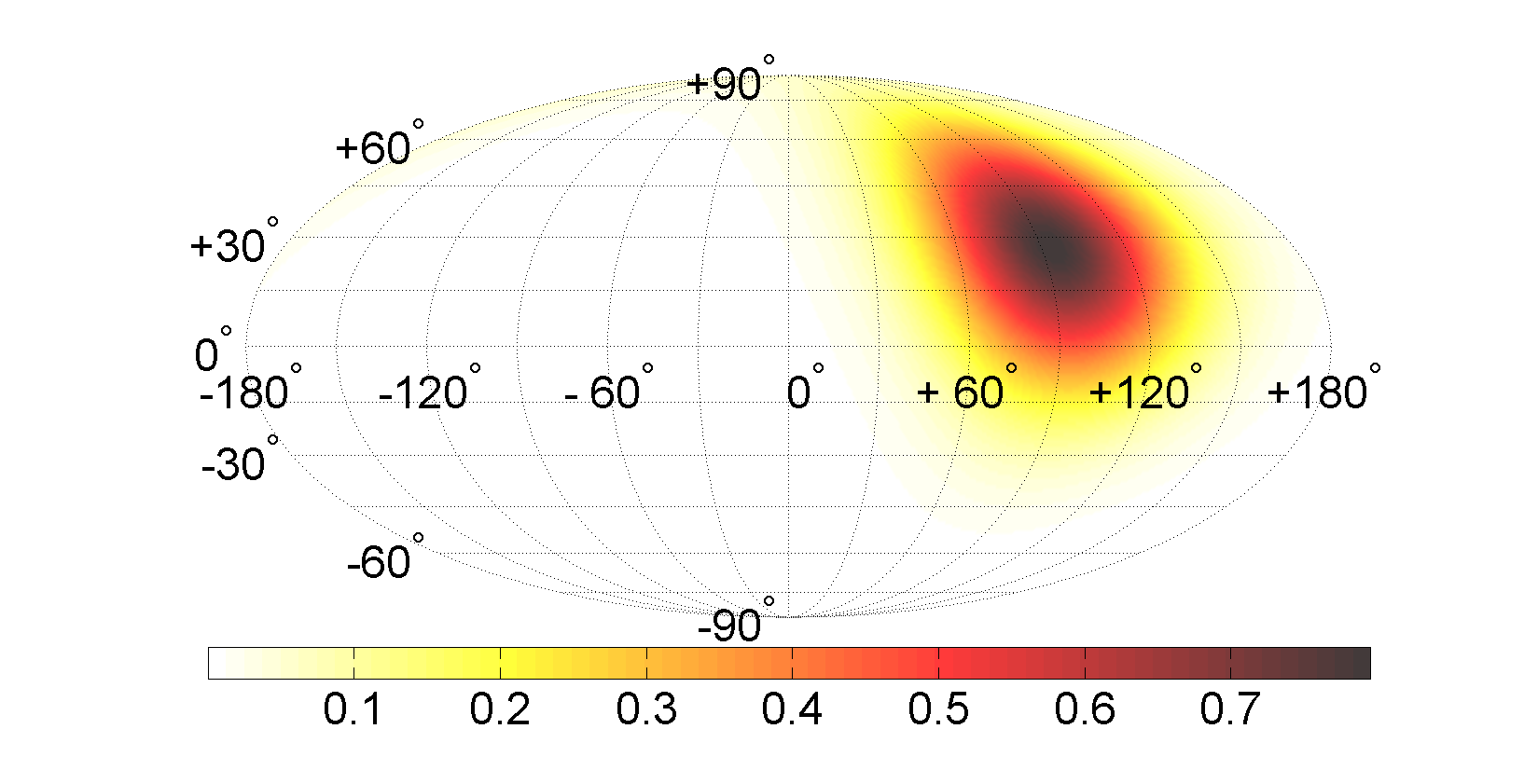}}%
\caption{}%
\label{fig:4b}%
\end{subfigure}~
\\
\begin{subfigure}{\columnwidth}
\centering
\scalebox{0.9}{\includegraphics[width=\textwidth]{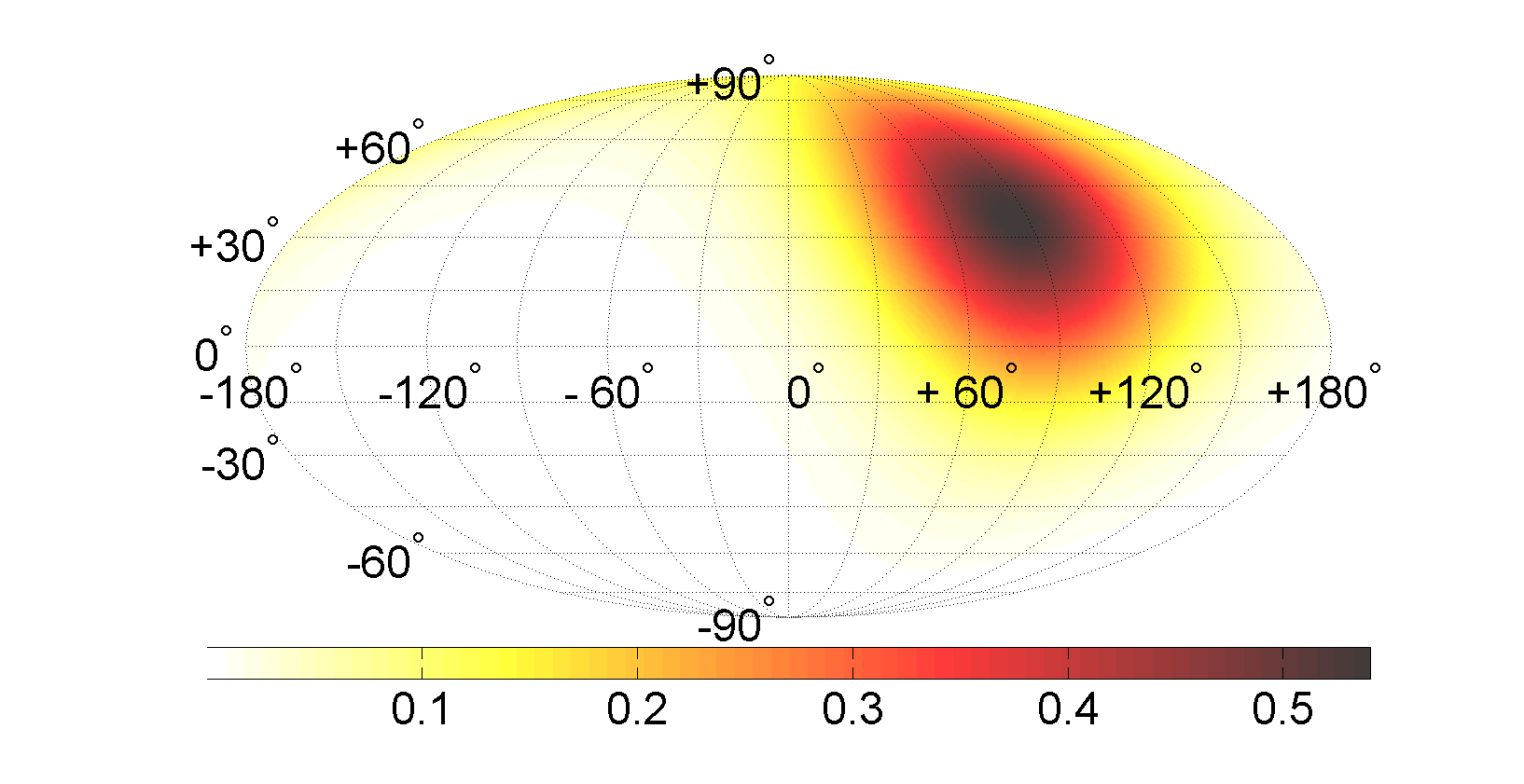}}%
\caption{}%
\label{fig:4c}%
\end{subfigure}~%
\vspace{-0.5em}
\\
\caption{3D von Mises-Fisher probability distribution function represented on a 2D Mollweide map projection for non-voice users, more specifically for phone usage types:  a) $1000$, b) $1010$, and c) $1100$. See Table.~\ref{tab:1} and Table.~\ref{tab:2} for phone usage type and distribution parameter specification, respectively.}
\vspace{-0.5em}
\label{fig:4}
\end{figure}

\section{Spherical model of the device orientation and validation procedure}\label{sec:3}
The actual orientation of a device is given by the $\theta$ and $\phi$ angles as shown in Fig.~\ref{figure:coordinate}. We propose here an equivalent way that we believe is more meaningful and useful.
\subsection{The von Mises-Fisher (vMF) directional distribution}
Indeed, from the physical considerations stated above we can see that the actual orientation of the device is given by the orientation of the gravity acceleration vector in the device local coordinate system $xyz$. In this case the tip of the ``random" vector $\boldsymbol{g}$ defined in this $xyz$-system will lay on the sphere with radius $g$. Hence, it is enough to define the vector indicating the orientation of the mobile phone device, i.e., an \emph{orientation vector}. For acceleration data sample $i$ we define it as
\begin{equation}\label{eq:2}
  \boldsymbol{\rho}_i=\frac{\boldsymbol{a}_i}{\|\boldsymbol{a}_i\|},
\end{equation}
where $\boldsymbol{\rho}_i=(\rho_{ix},\rho_{iy},\rho_{iz})$ with $\|\boldsymbol{\rho}_i\|=1$.

A useful model describing the statistical distribution of a unit vector $\boldsymbol{\rho} \in \mathbb{R}^3$ is the von Mises-Fisher (vMF) probability distribution function (pdf) \cite{mardia2009directional}
\begin{equation}\label{eq:3}
  f(\boldsymbol{\rho};\boldsymbol{\mu},\kappa)=\frac{\kappa}{4\pi\sinh(\kappa)}\exp(\kappa \boldsymbol{\mu}^{\mathrm{T}}\boldsymbol{\rho}),
\end{equation}
where the parameter $\boldsymbol{\mu}$ with $\|\boldsymbol{\mu}\|=1$, is the \emph{mean direction} and $\kappa>0$, is the \emph{concentration parameter}. The distribution is rotationally symmetric around the mean direction $\boldsymbol{\mu}$, which provides the direction of concentration of data. The higher the spread of the data, the lower is $\kappa$, and viceversa. For uniformly distributed points on the sphere (i.e., isotropically distributed), $\kappa=0$ and \eqref{eq:3} becomes $f(\boldsymbol{\rho})=\frac{1}{4\pi}$; no mean direction can be defined in this case. The maximum-likelihood estimators of the two parameters $\boldsymbol{\hat{\mu}}$ and $\hat{\kappa}$ are known and are computed following the results provided in \cite{sra2012short}. For the sake of completeness of exposition the estimators are given in appendix A.

There are known algorithms to generate random vectors $\boldsymbol{\rho}$ following the distribution \eqref{eq:3} \cite{wood94}. In our computations we use the MATLAB$^{\circledR}$ script provided in \cite{vmfgener}.
\begin{figure}[!ht]%
\centering
\begin{subfigure}{\columnwidth}
\centering
\scalebox{0.9}{\includegraphics[width=\textwidth]{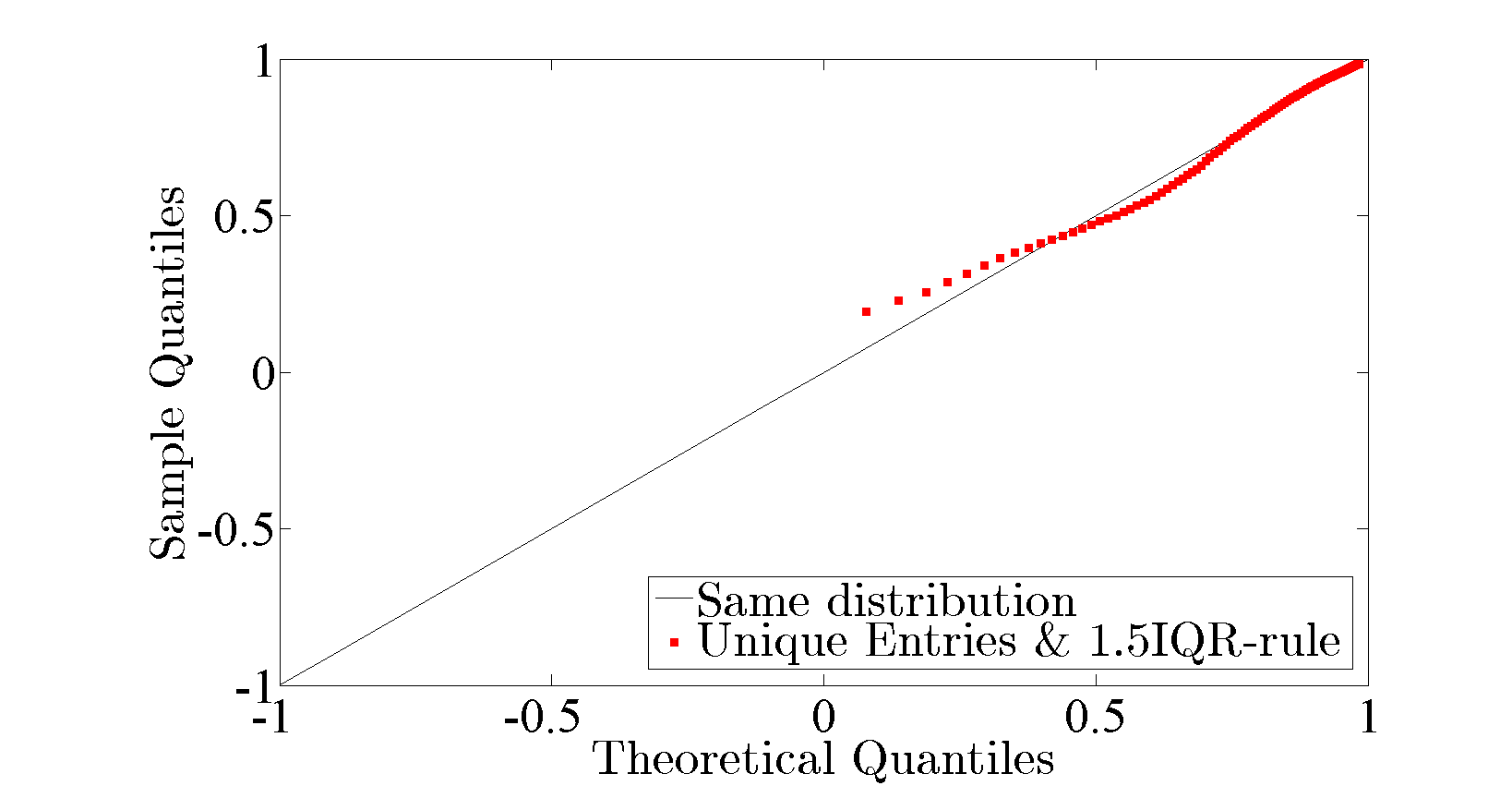}}%
\caption{}%
\label{fig:5a}%
\end{subfigure}~
\\
\begin{subfigure}{\columnwidth}
\centering
\scalebox{0.9}{\includegraphics[width=\textwidth]{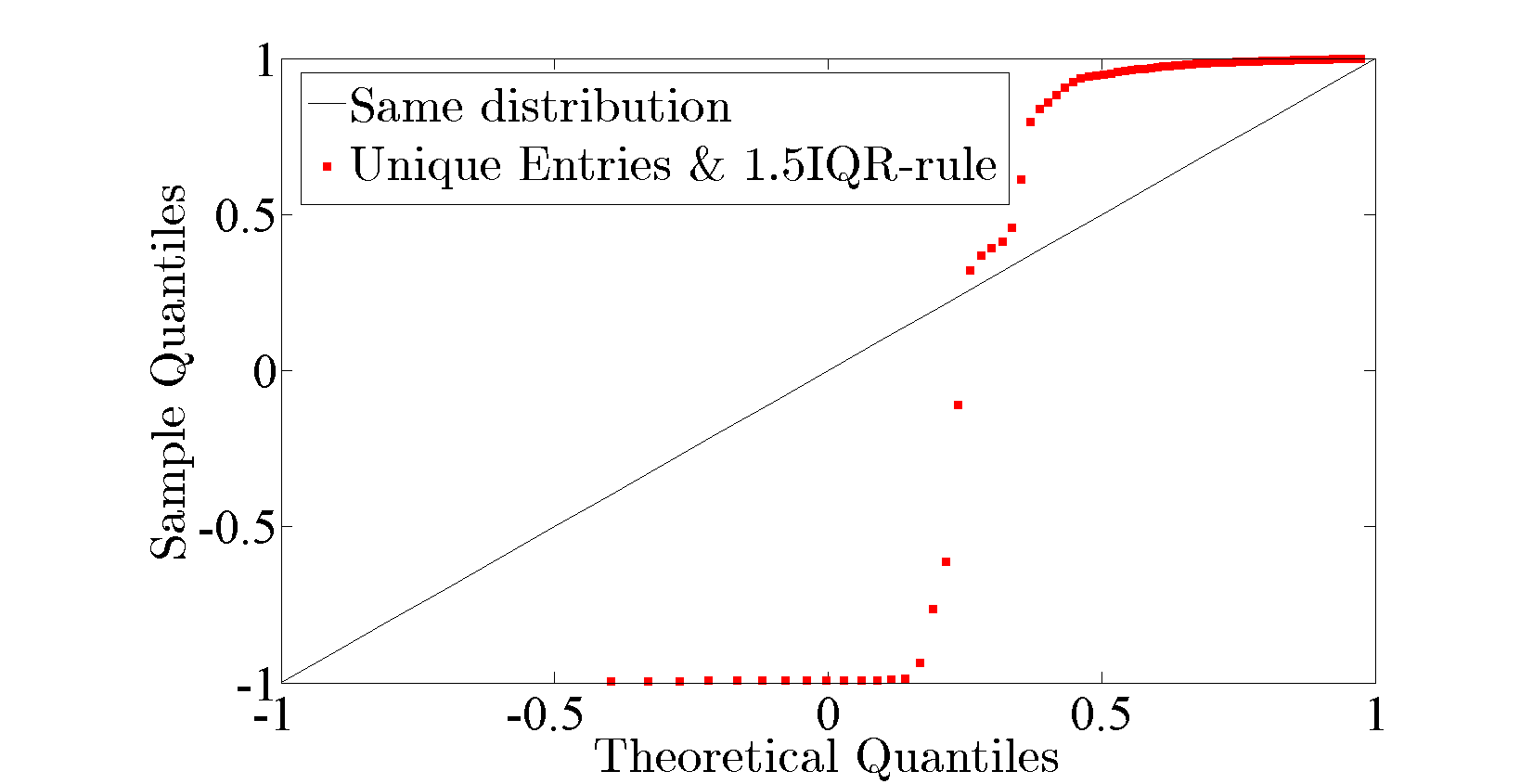}}%
\caption{}%
\label{fig:5b}%
\end{subfigure}~
\\
\begin{subfigure}{\columnwidth}
\centering
\scalebox{0.9}{\includegraphics[width=\textwidth]{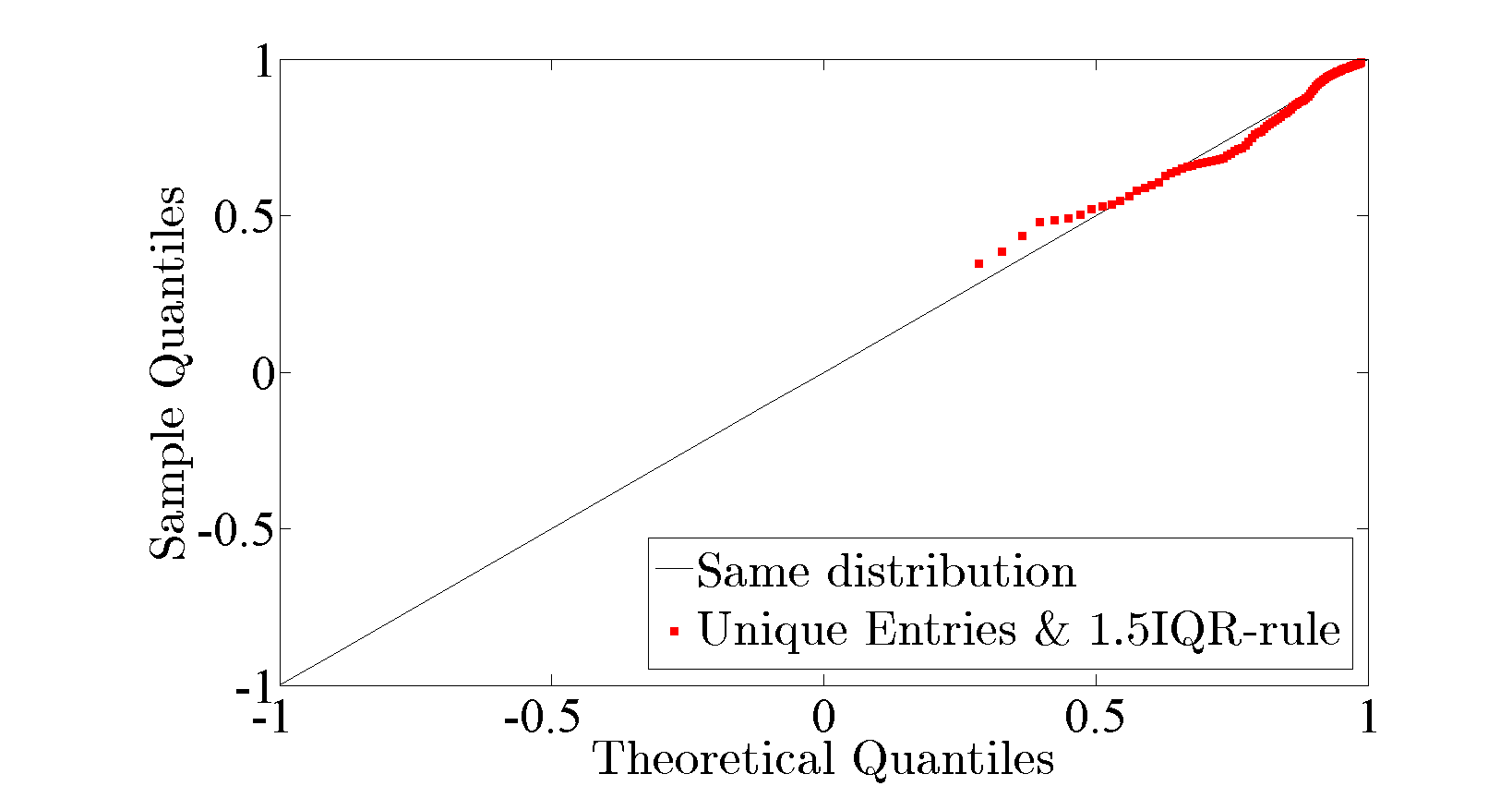}}%
\caption{}%
\label{fig:5c}%
\end{subfigure}~%
\\
\begin{subfigure}{\columnwidth}
\centering
\scalebox{0.9}{\includegraphics[width=\textwidth]{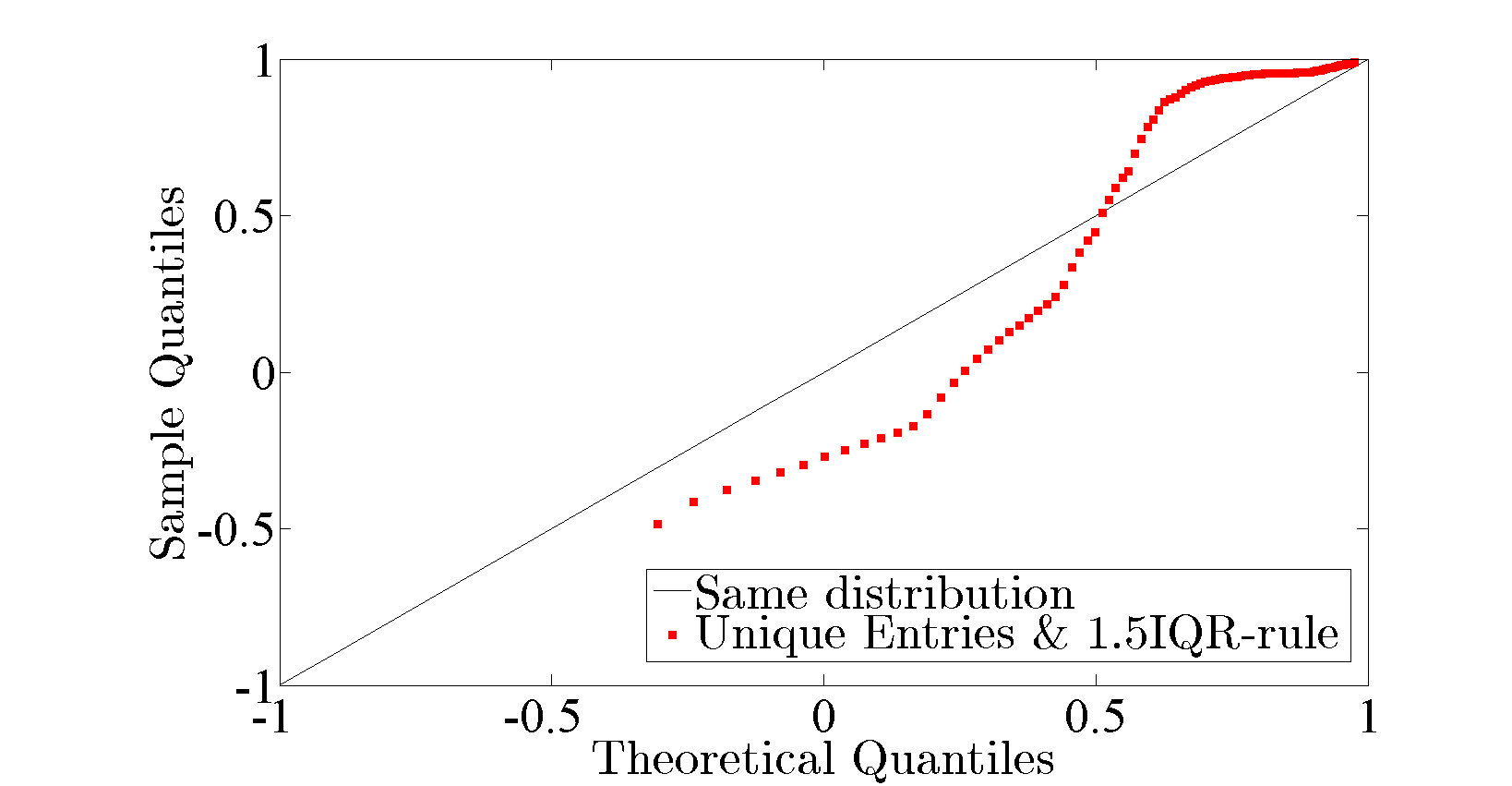}}%
\caption{}%
\label{fig:5d}%
\end{subfigure}%
\\
\vspace{0.5em}
\caption{Q-Q plots for voice users, more specifically for phone usage types;  a) $0000$, b) $0001$, c) $0010$ and d) $0100$. See Table.~\ref{tab:1} and Table.~\ref{tab:2} for phone usage type and distribution parameter specification, respectively.}
\label{fig:5}
\end{figure}

\begin{figure}[!ht]%
\centering
\begin{subfigure}{\columnwidth}
\centering
\scalebox{0.9}{\includegraphics[width=\textwidth]{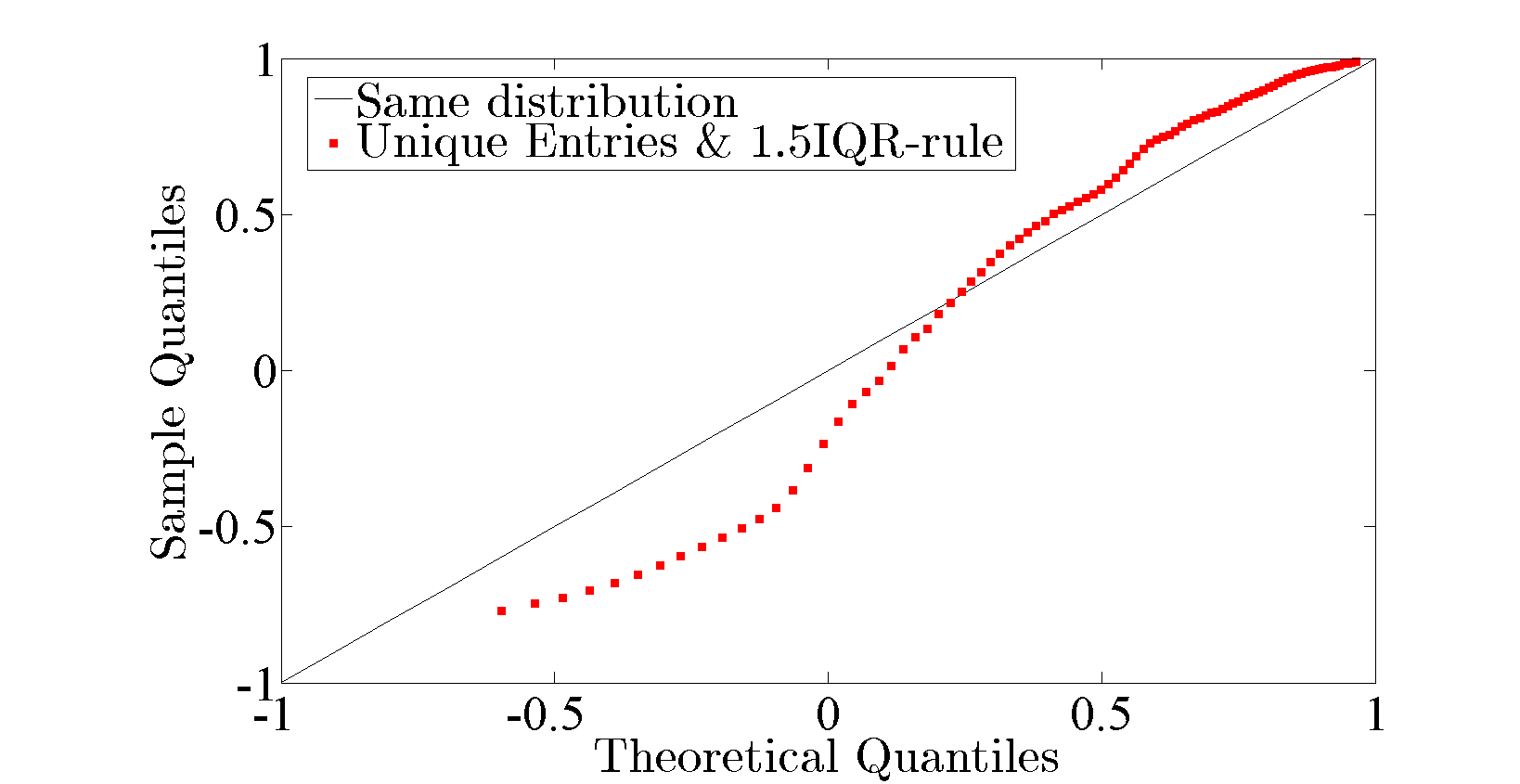}}%
\caption{}%
\label{fig:6a}%
\end{subfigure}~
\\
\begin{subfigure}{\columnwidth}
\centering
\scalebox{0.9}{\includegraphics[width=\textwidth]{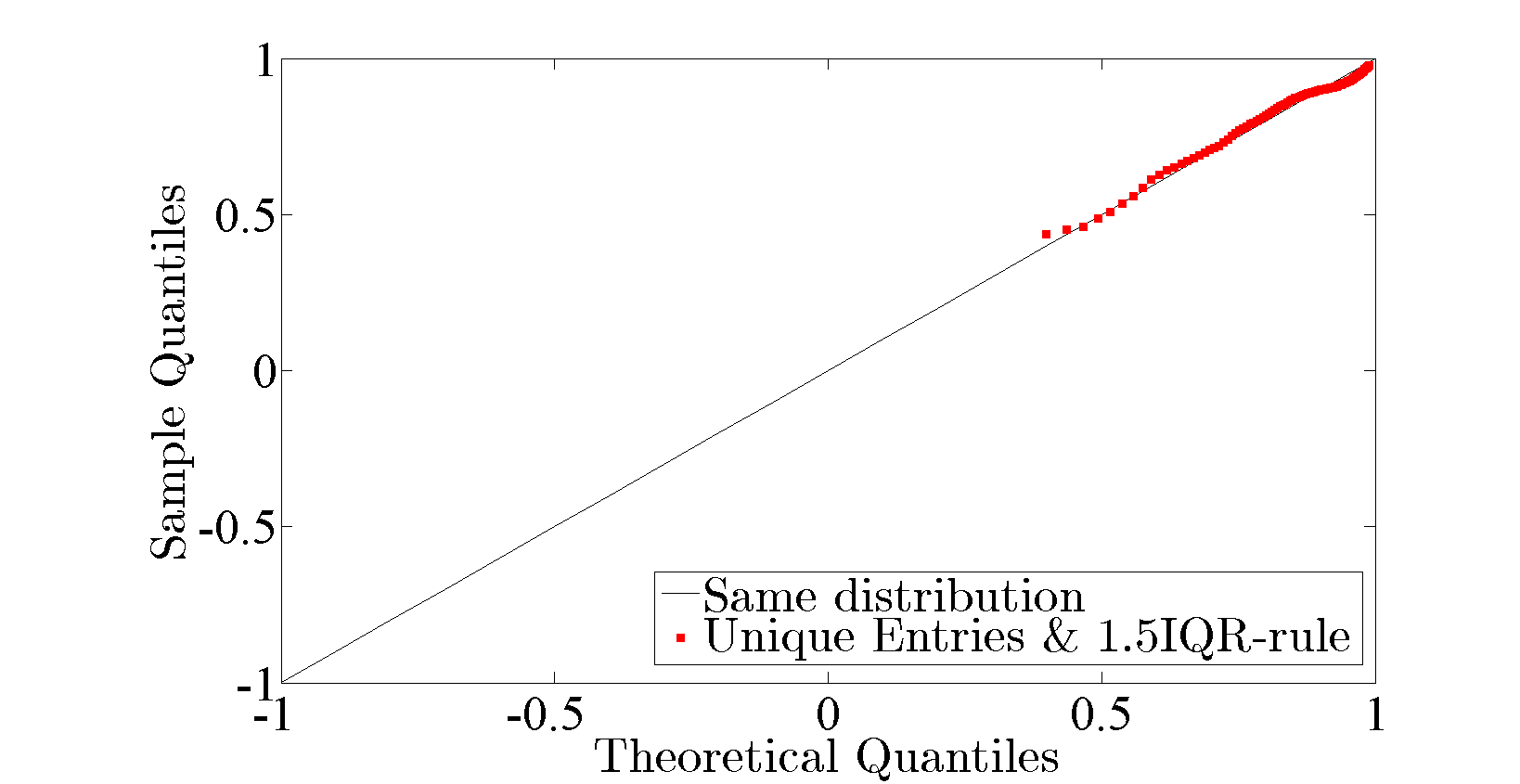}}%
\caption{}%
\label{fig:6b}%
\end{subfigure}~
\\
\begin{subfigure}{\columnwidth}
\centering
\scalebox{0.9}{\includegraphics[width=\textwidth]{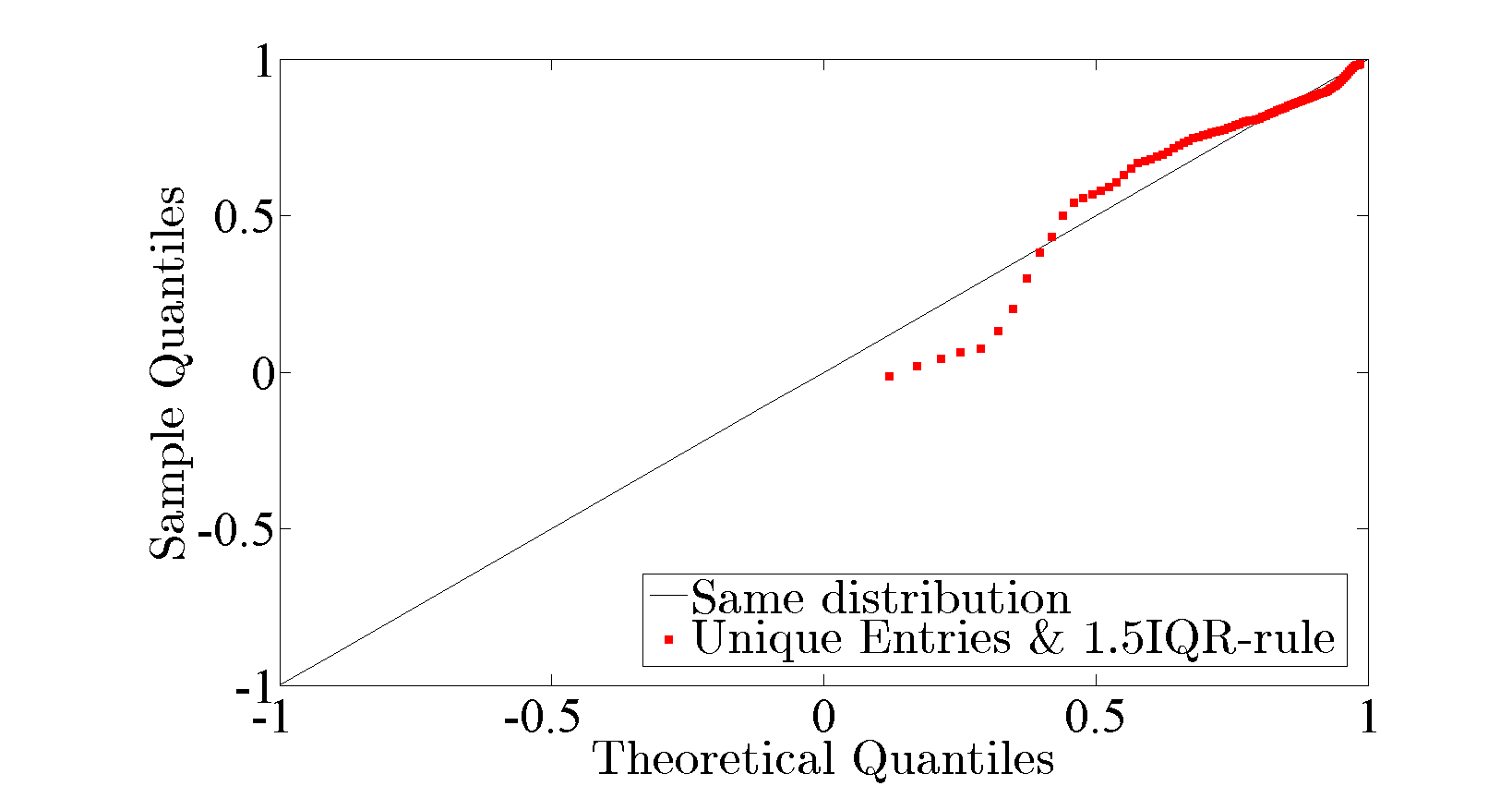}}%
\caption{}%
\label{fig:6c}%
\end{subfigure}~%
\vspace{0.5em}
\caption{Q-Q plots for non-voice users, more specifically for phone usage types;  a) $1000$, b) $1010$, and c) $1100$. See Table.~\ref{tab:1} and Table.~\ref{tab:2} for phone usage type and distribution parameter specification, respectively.}
\vspace{-0.5em}
\label{fig:6}
\end{figure}
\subsection{Q-Q plots for the vMF directional distribution}
The vMF distribution parameters, $\boldsymbol{\hat{\mu}}$ and $\hat{\kappa}$, i.e., the estimates of the mean direction and the concentration parameter, respectively, fully define our model of the directional data. However, we need to assess how well the model fits our data. An illustrative way is just plotting the normalized acceleration vectors or orientation vectors \eqref{eq:2}. However, given the large number of measurement points this is not a very practical way. Besides, we need a more meaningful approach. In this paper we use the so-called Quantile-Quantile plot or Q-Q plot for short. A Q-Q plot is a graphical method showing the quantiles of one probability distribution against the quantiles of another probability distribution. Hence, two probability distributions can be definitely compared and if two data samples come from the same probability distribution or data comes from the same theoretical probability distribution as the model then the Q-Q plot is a straight line. It is worthwhile to note that the Q-Q plot only gives an approximate idea of the underlying distribution.

In our analysis we use the Q-Q plot concept for directional statistics introduced in \cite{ley2014}. It is straightforward to implement, to interpret and it has many other advantages therein. The applicability of the method, i.e., the new quantile definition is restricted to the class of directional probability distributions with bounded density, admit a unique median direction, and are rotationally symmetric about the mean direction, which are satisfied by the vMF distribution. Due to practical reasons, we consider in this paper the empirical version for computing the \emph{projection quantile}, i.e., $\hat{c}_{\tau}$. For a given quantile $\tau \in [0, 1]$, the corresponding projection $\tau-$quantile is obtained in three steps: (1) estimate the median $\boldsymbol{\mu}$ by an estimator $\boldsymbol{\hat{\mu}}$, (2) project all observations onto $\boldsymbol{\hat{\mu}}$, i.e., the projected population is $\boldsymbol{\rho}^{\mathrm{T}}\boldsymbol{\hat{\mu}}$ and (3) use a traditional definition of univariate quantiles for determining the $\hat{c}_{\tau}$.
\section{Data processing, model fitting results and analysis}\label{sec:4}
The analysis includes data that has been collected from $11$ users which used $7$ different smart phone models from $4$ different brands. Disclosing further details is outside the scope of this paper since we focus on the orientation statistics.
\subsection{Data pre-processing}
Prior to proceeding with the model fitting, the accelerometer data was extracted following the binary categories given in Table.~\ref{tab:1}. Then, outliers were removed following the \emph{$1.5$IQR-rule} and duplicate data entries were removed by taking unique values. The same procedure was applied to all phone usage types that could be extracted from the collected data. The phone usage types are listed in the first column in Table.~\ref{tab:2}.

An example of the results of the data processing is given in Fig.~\ref{fig:2}. The upper subplot shows the Cumulative Distribution Function (CDF) of the magnitude of the acceleration $\|\boldsymbol{a}\|$ corresponding to the voice service for users that used no wired-headset, no speaker-phone and no Bluetooth-headset. Results are presented for the raw data as downloaded from the app and data that has been processed as explained above. As expected, the applied outlier elimination rule and selecting unique values moved the distribution of $\|\boldsymbol{a}\|$ toward the median, i.e., the gravity acceleration $g$.

The three lower subplots in Fig.~\ref{fig:2} show the Probability Distribution Function (PDF), or rather, the normalized histogram of $\rho_{ix}$, $\rho_{iy}$, $\rho_{iz}$ components of the corresponding normalized acceleration vector, i.e., the phone orientation vector \eqref{eq:1}. As can be seen the data processing efficiently removed as turned out to be duplicated values in the vicinity of $\rho_{ix}=\rho_{iy}=0$ and for $\rho_{iz}=1$. It is worthwhile to recall that corresponding acceleration values where removed prior normalization. Hence, we expect the processed data be more representative of the user-induced randomness due to device orientation. The presented approach should be considered as a straightforward, yet approximate, identification procedure of useful random data samples.

\begin{figure}[!t]
	\centering
	\begin{subfigure}{\columnwidth}
		\centering
		\scalebox{0.9}{\includegraphics[width=\textwidth]{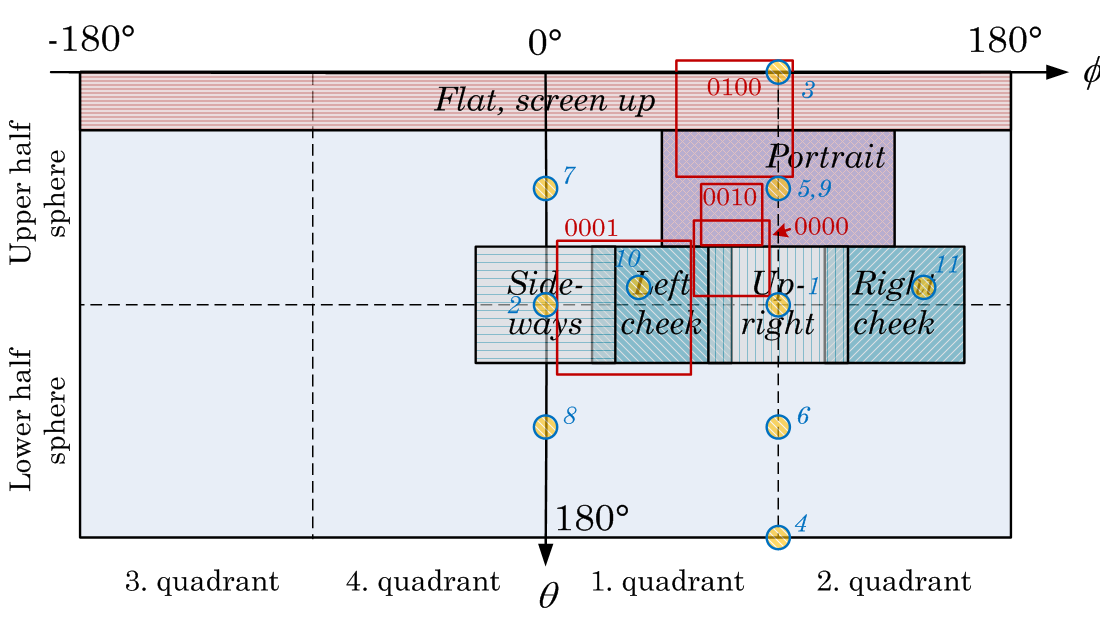}}%
		\caption{}%
		\label{fig:7a}%
	\end{subfigure}~
	\\
	\begin{subfigure}{\columnwidth}
		\centering
		\scalebox{0.9}{\includegraphics[width=\textwidth]{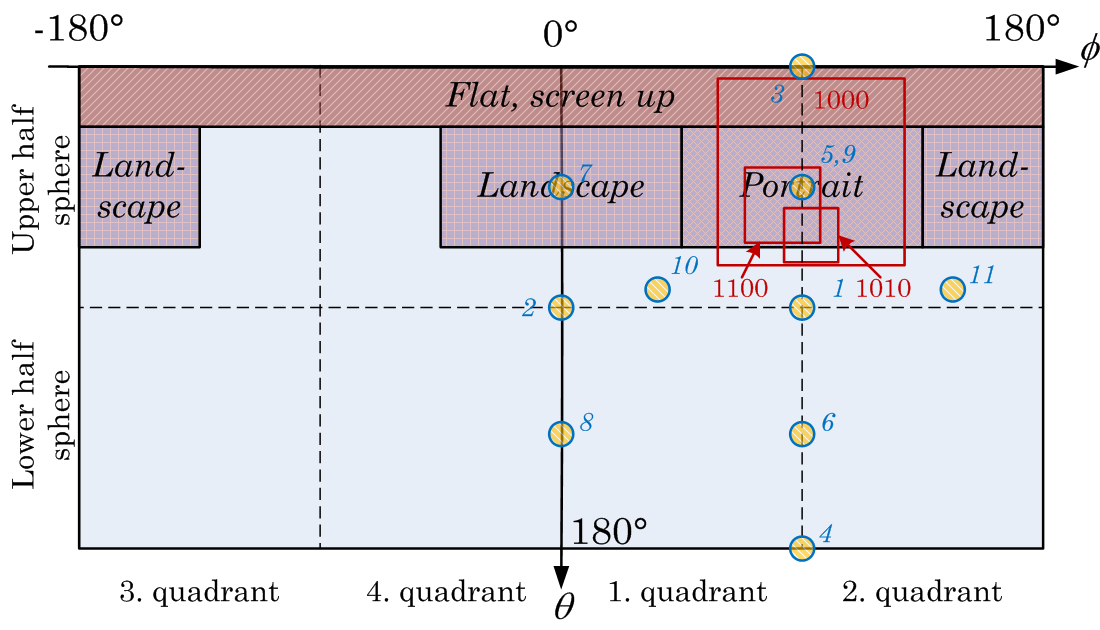}}%
		\caption{}%
		\label{fig:7b}%
	\end{subfigure}~
	\\
	\vspace{0.5em}
	\caption{$\phi$-$\theta$ plots showing the location and spreading of the different usage types. Size of squares are inversely proportional to the concentration factor $\kappa$;  a) voice b) non-voice. See Table.~\ref{tab:1} and Table.~\ref{tab:2} for phone usage type and distribution parameter specification, respectively. The circles show the OTA test positions as defined by the 3GPP and explained in Table.~\ref{tab:3}.}
	\label{fig:7}
\end{figure}

\begin{table*}[!t]
	\renewcommand{\arraystretch}{1.3}
	\caption{OTA test conditions as defined by the 3GPP \cite{3GPPTR37977}.}
	\centering
	\setlength{\tabcolsep}{3pt}
	\begin{tabular}{c|l|l}
		\hline \hline
		Index & 3GPP test condition & Comment \\
		\hline \hline
		1 & XY-plane & Vertical upright \\
		2 & XZ-plane & Vertical sideways \\
		3 & Free space data mode (FS-DMSU) & Horizontal, screen up \\
		4 & Face down & Horizontal. screen down \\
		5 & Free space data mode portrait (FS-DMP) & Portrait, tilted \\
		6 & Free space tilted down & Portrait, downtilted \\
		7 & Free space data mode landscape (FS-DML) & Landscape, tilted \\
		8 & Free space landscape, tilted down & Landscape, downtilted \\
		9 & Left/right hand phantom data mode portrait (LH/RH-DMP) & Portrait, tilted \\
		10,11 & Beside head and hand right/left (BHHR/BHHL) & Cheek right and cheek left \\
		\hline \hline
	\end{tabular}\label{tab:3}
	
\end{table*}

\begin{table*}[!t]
	\renewcommand{\arraystretch}{1.3}
	\caption{Matching phone usage types and CTIA/3GPP test conditions.}
	\centering
	\setlength{\tabcolsep}{3pt}
	\begin{tabular}{c|c|l}
		\hline \hline
		Ph. Usage $\{ijkl\}$ & 3GPP Test Cond. & Comment \\
		\hline \hline
		0000 & 1 & Voice, no wired or wireless headset. Vertical upright \\
		0001 & 10 & Voice, Bluetooth on. Vertical slant \\
		0010 & 5,9 & Voice, speakerphone. Portrait, tilted \\
		0100 & 3 & Voice, wired headset. Horizontal screen up \\
		\hline \hline
		1000 & 5,9 & Non-voice, no wired or wireless headset. Portrait, tilted \\
		1010 & 1,5,9 & Non-voice, speakerphone. Portrait, vertical to tilted \\
		1100 & 5,9 & Non-voice, wired headset. Portrait, tilted \\
		\hline \hline
	\end{tabular}
\end{table*}\label{tab:4}

\subsection{von Mises-Fisher distribution fitting}
\label{sub:vMF-fitting}
Fig.~\ref{fig:3} and Fig.~\ref{fig:4} show the 3D vMF directional PDFs represented on a 2D Mollweide map projection corresponding to voice and non-voice phone usage types, respectively. The functions where generated using \eqref{eq:3} and using the distribution parameters estimates (see Appendix A) given in Table.~\ref{tab:2}. The presented plots provide a full spherical coverage of the statistical distribution for both voice and non-voice phone usage. We can clearly see the main direction for each usage type, i.e., the mean phone orientation in the different usage orientations. As explained above, the lower the $\kappa$ parameters the less concentrated values of the orientations around the mean orientation and viceversa.

Fig.~\ref{fig:5} and Fig.~\ref{fig:6} show the Q-Q plots corresponding to sample vMF distributions v.s. theoretical (simulated) vMF distributions computed with estimated parameters given in Table.~\ref{tab:2}. Each plot in Fig.~\ref{fig:5} and Fig.~\ref{fig:6} should be considered in juxtaposition with corresponding plots in Fig.~\ref{fig:3} and Fig.~\ref{fig:4}, respectively. The estimated projection quantiles $\hat{c}_{\tau}$ where obtained for quantiles $\tau=\{0.05,... ,0.95\}$ taken in $0.01$ steps. Hence, the leftmost point corresponds to $\tau=0.05$ ($5\%$-quantile), while the rightmost point corresponds to $\tau=0.95$ ($95\%$-quantile). By visual inspection we immediately realize that the proposed vMF model only approximately describes the directional distribution of the data. The best fits correspond to phone usage type $0000$, $0010$ and $1010$, followed by less so models corresponding to phone usage type $1000$, $1100$ and $0100$ and the worst fitting is for phone usage type $0001$.

The main reason that may lead to discrepancies between the model and the data samples is the fact that the underlying true empirical distribution is not well-described by a single cluster. Therefore, future models should take into account multiple clusters and a possible correlation between them. However, the presented results may be considered as a first-order approximation describing the user-induced mobile phone orientation randomness. And as shown above, for some specific cases, the vMF is indeed an excellent fit to the statistical distribution of the orientation of the mobile phone.
\subsection{Finite mixture model for user induced randomness}
The vMF distribution is indeed a straightforward model that together with parameters provided in Table.~\ref{tab:2} gives an easy way to simulate different phone orientations following the algorithm in \cite{wood94}; also the MATLAB scripts available in \cite{vmfgener} can be used. However, we need more, we need a model with as much flexibility as needed to accurately describe the real life data.

The concept of a finite mixture model (FMM) provides us with the desired model flexibility \cite{mclachlan2004finite}. We may define an FMM of vMF distributions describing different phone usage type as the device orientation vector $\boldsymbol{\rho}$ with the PDF
\begin{align}\label{eq:4}
f_\mathrm{mix}(\boldsymbol{\rho};\boldsymbol{\gamma})&=\frac{\sum_{i=0}^1\sum_{j=0}^1\sum_{k=0}^1\sum_{l=0}^1\pi_{ijkl}f(\boldsymbol{\rho};\boldsymbol{\mu}_{ijkl},\kappa_{ijkl})}{\sum_{i=0}^1\sum_{j=0}^1\sum_{k=0}^1\sum_{l=0}^1\pi_{ijkl}},
\end{align}
where $f(\boldsymbol{\rho};\boldsymbol{\mu}_{ijkl},\kappa_{ijkl})$ is the vMF distribution with mean direction $\boldsymbol{\mu}_{ijkl}$ and concentration parameters $\kappa_{ijkl}$ and the weights $\pi_{ijkl}\geq 0$ satisfy the probability normalization
\begin{align}
\sum_{i=0}^1\sum_{j=0}^1\sum_{k=0}^1\sum_{l=0}^1\pi_{ijkl}&=1.  \label{eq:5}
\end{align}
The parameter
\begin{align}
\boldsymbol{\gamma}&=\{\boldsymbol{\mu}_{ijkl},\kappa_{ijkl},\pi_{ijkl}\}~\text{with}~i,j,k,l=\{0,1\}, \label{eq:6}
\end{align}
denotes the parameter matrix of the FMM. The number of components is itself a parameter of the model.

The drawback is that estimating the parameters of the vMF FMM is not a trivial task and requires a rather sophisticated approach to solving the task at hand \cite{mclachlan2004finite}. Here, we propose a heuristic approach exploiting the knowledge about the data that we already have. Indeed, since we define the phone usage types,  we can then in turn compute the frequency of observing data that correspond to a certain phone usage type $\pi_{ijkl}$ as
\begin{align}
\pi_{ijkl}&=\frac{N_{\mathrm{s},ijkl}}{\sum_{i=0}^1\sum_{j=0}^1\sum_{k=0}^1\sum_{l=0}^1 N_{\mathrm{s},ijkl}}, \label{eq:7}
\end{align}
where $N_{\mathrm{s},ijkl}$ denotes the number of samples used to estimate the vMF distribution parameters. Clearly, \eqref{eq:7} satisfies the normalization \eqref{eq:5}. Assuming a vMF FMM that includes all the practical phone usage types, we can then obtain the heuristic weights as given in Table.~\ref{tab:2}. Terms with missing values are just replaced by $0$ in the summations in \eqref{eq:4}, \eqref{eq:5} and \eqref{eq:7}. In this way we have now computed all the values of the FMM parameter matrix $\boldsymbol{\gamma}$ in \eqref{eq:6}. The PDF of the vMF FMM  is now fully defined, which can be straightforwardly simulated following standard algorithms, e.g., the composition method algorithm \cite{kroese2013handbook}.

In addition to the above model, other mixture models can be generated with the same procedure when applied to subsets of phone usage types according to Table.~\ref{tab:2}. For example, it may be interesting to consider usage types when no headsets nor speaker phone are used. In that case the vMF FMM distribution is computed by with the help of the parameters corresponding to phone usage types $0000$ and $1000$ in Table.~\ref{tab:2}. It is worthwhile to note that in this case the weights $\pi_{0000}$ and $\pi_{1000}$ do not sum up to $1$. However, the formulation of the vMF FMM \eqref{eq:4} guaranties that the renormalized weights $\frac{\pi_{0000}}{\pi_{0000}+\pi_{1000}}$ and $\frac{\pi_{1000}}{\pi_{0000}+\pi_{1000}}$ do sum up to $1$ as required.

Hence, differentiating the most common phone usages for the two main services enables a systematic modelling of the device orientation, with applications to OTA testing, to wireless network analysis and optimization, to radio wave propagation and indoor localization.

\subsection{Mapping phone usage types to typical usage positions}
In \cite{UserRandomNess_Lehne_16}, the concept of \emph{user modes} was introduced. The hypothesis is that certain ways of holding the phone should dominate depending on the services accessed. For voice usage it is expected that use of handsfree, whether wired, bluetooth or loud-speaker, would be different than without. We see from the vMF distribution parameters in Table~\ref{tab:2} that this is the case. Traditional usage for voice calls would be holding the phone to either right or left cheek, while handsfree usage could include a number of modes. For non-voice usage, we could expect that some kind of slant holding positions, either portrait or landscape mode would be typical, but also having the phone lying horizontally on a flat surface.

Industry and standards organizations CTIA and 3GPP has developed test plans for OTA performance measurements of user equipment \cite{3GPPTR37977, CTIAMimoOta}. Here, fixed orientation conditions for Equipment Under Test (EUT) has been defined, including a number of EUT orientation angles. These correspond to our definition of user modes, and some anticipated modes are drawn in the $\phi$-$\theta$ coordinate system in Fig.~\ref{fig:7} together with the found statistics from Table~\ref{tab:2}. The user modes corresponding to the CTIA and 3GPP orientation angles are shown in the same figures and explained in~\ref{tab:3}. For voice usage, we see that an upright position tending towards left cheek is common without handsfree (usage type $0000$), while handsfree usage results in more slant positions. Use of wired handsfree is near to horizontal flat mode. For non-voice usage, most usage types maps to portrait orientation, except the use of speaker which shows a vertical position.

The statistics analysis in \cite{UserRandomNess_Lehne_16} showed that there are multiple clusters, especially for usage type $0000$, and that is missed here, as discussed in subsection~\ref{sub:vMF-fitting}. Two quite visible peaks on left cheek and right cheek was visible, where the left position was most common. The preference for left cheek is only visible by the centre position of the $0000$ usage type.

The CTIA and 3GPP defined orientation angles are not given as distributions, but as fixed points. As can be seen from Tables \ref{tab:3} and \ref{tab:4}, only some of the definitions seem relevant. For example, for voice usage this are $3$, $5/9$ and $10$, while for non-voice usage it is basically $5/9$. Orientations corresponding to test conditions $2$, $4$, $6-8$ and $11$ are not observed. It is worthwhile to note that the model is not handling distributions with multiple maxima.
\section{Conclusions and future work}\label{sec:5}
In this paper we have shown how smart-phone sensors can be used to model a wireless device usage in real-life situations without the intervention of the experimenter. This information can have a profound repercussion on the design of better handsets, but also to improve wireless network performance.

Indeed, the proposed von Mises-Fisher directional distribution provides a statistical model for the phone's orientation over the full sphere for different \emph{phone usage types}, e.g., voice or non-voice services used in combination with wired or Bluetooth handsfree connection or a speaker. Also different mixes of phone usage types can be studied and simulated based on their individual distributions by the finite mixture model approach. The heuristic weights of the mixture model are computed as the frequency of observation of data samples belonging to the different phone usage types. Hence, statistical distributions describing phone usage types in voice and non-voice modes can be generated and incorporated, e.g., in Random-LOS OTA performance analysis, channel model generation and wireless network design and optimization.

The model is based on analysing the acceleration (gravity) vector data obtained from accelerometer sensor readings in real life. However, the analysis of the data collected from the users may not be fully representative of the actual user-induced mobile phone orientation randomness since data is limited to $11$ users so far. However, the number of users reporting their accelerometer data will increase in the future, e.g., by means of crowdsourcing. This will provide a more representative data set of the phone usage types and therefore improve the randomness characterization of the users. Also, in future analysis more sophisticated data pre-processing (i.e., before the statistical parameters are estimated) will be needed in order to fine-tune the statistical models of the different phone usage types. For example, the finite mixture model could be applied to each usage type separately to identify clustering effects that might have been gone unnoticed by removing duplicate data, i.e., the pre-processing approach we have used in our analysis.

\section*{Appendix A}
Let $\chi_{\boldsymbol{\rho}} = \{\boldsymbol{\rho}_1,\ldots, \boldsymbol{\rho}_N\}$ be a set of points drawn from $f(\boldsymbol{\rho};\boldsymbol{\mu},\kappa)$ \eqref{eq:3}. The estimates $\boldsymbol{\hat{\mu}}$ and $\hat{\kappa}$ that maximize the log-likelihood function
\begin{equation} \label{eq:A1}\\
\mathcal{L}(\chi_{\boldsymbol{\rho}};\boldsymbol{\mu},\kappa)=N\ln\left(\frac{\kappa}{4\pi\sinh(\kappa)}\right)+\sum_{i=1}^N \kappa \boldsymbol{\mu}^{\mathrm{T}}\boldsymbol{\rho}_i,
\end{equation}
subject to the condition that $\boldsymbol{\mu}^{\mathrm{T}}\boldsymbol{\mu} = 1$ and $\kappa \geq 0$ are given by \cite{sra2012short}
\begin{align}
\boldsymbol{\hat{\mu}} &=\frac{\sum_{i=1}^N\boldsymbol{\rho}_i}{\| \sum_{i=1}^N\boldsymbol{\rho}_i \|}, \label{eq:A2}\\
\hat{\kappa} &=A_3^{-1}(\bar{R}), \label{eq:A3}
\end{align}
where
\begin{align}
\bar{R}&=N^{-1}\| \sum_{i=1}^N\boldsymbol{\rho}_i \|, \label{eq:A4}\\
A_3(\kappa)&=I_{3/2}(\kappa)/I_{1/2}(\kappa), \label{eq:A5}
\end{align}
where $I_{\nu}$ denotes the modified Bessel function of the first kind of order $\nu$ \cite{Gradshteyn2000}. Hence, $\hat{\kappa}$ is the solution to $A_3(\kappa)-\bar{R}=0$, that is obtained numerically by the Newton's method
\begin{align}
\kappa_0&=\frac{\bar{R}(3-\bar{R}^2)}{1-\bar{R}}, \label{eq:A6}\\
\kappa_{k}&=\kappa_{k-1}-\frac{A_3(\kappa_{k-1})-\bar{R}}{1-A_3(\kappa_{k-1})^2-\frac{2}{\kappa_{k-1}}A_3(\kappa_{k-1})}, \label{eq:A7}
\end{align}
where $\kappa_0$ is the initial value and the iteration usually converges after two steps, and thus $\hat{\kappa}=\kappa_2$.

\bibliographystyle{IEEEtran}

\end{document}